\documentclass[12pt]{article}
\usepackage{amsmath,amssymb,amsthm,amscd}
\usepackage{jheppub}
\usepackage{verbatim}
\usepackage{amsfonts}

\usepackage{tikz}

\usetikzlibrary{decorations.markings,arrows,backgrounds,patterns,decorations.pathreplacing}

\tikzset{>=latex}
\tikzset{every picture/.append style={scale=1.1}}
\tikzset{flavor/.style={draw}}
\tikzset{gauge/.style={draw,circle,inner sep=2pt}}

\allowdisplaybreaks

\def\Z{{\cal Z}}
\def\be{\begin{equation}}
\def\ee{\end{equation}}
\def\ba{\begin{aligned}}
\def\ea{\end{aligned}}
\def\ben{\begin{eqnarray}\displaystyle}
\def\een{\end{eqnarray}}

\def\CT{{\cal T}}
\def\CI{{\cal I}}
\def\CN{{\cal N}}

\def\IC{{\mathbb C}}
\def\IR{{\mathbb R}}
\def\IZ{{\mathbb Z}}

\def\I{{\mathfrak{I}}}

\def\mt{{m}}

\def\qt{{q}}

\def\nn{\nonumber}

\def\a{{\alpha}}

\def\a{{\alpha}}
\def\tm{{m}}

\newcommand{\q}{\mathbf{q}}
\newcommand{\Qi}{{\tt q_1}}
\newcommand{\Qii}{{\tt q_2}}
\newcommand{\Qiii}{{\tt q_3}}
\newcommand{\QI}{{\tt q_i}}

\title{Symmetry enhancements via 5d instantons, $q\mathcal{W}$-algebrae and $(1,0)$ superconformal index}

\author{Sergio Benvenuti, Giulio Bonelli, Massimiliano Ronzani and Alessandro Tanzini}
\affiliation{SISSA/ISAS, via Bonomea 265, 34136, Trieste, Italy and INFN, Sezione di Trieste}
\emailAdd{sbenvenu,bonelli,mronzani,tanzini@sissa.it}

\abstract{We explore $\mathcal{N}=(1,0)$ superconformal six-dimensional theories arising from M5 branes probing a transverse $A_k$
singularity. Upon circle compactification to 5 dimensions, we describe this system
with a dual pq-web of five-branes and propose the spectrum of basic five-dimensional instanton operators driving global symmetry enhancement. For a single
M5 brane, we find that the exact partition function of the 5d quiver gauge theory matches the 6d $(1,0)$ index, which we compute by letter counting.
We finally show that S-duality of the pq-web implies new relations among vertex correlators of $q\mathcal{W}$ algebrae.}

\begin{document}
\maketitle

\section{Introduction and results}

The description of systems of multiple M5 branes is still an elusive problem in our current understanding of M-theory. Nonetheless, many progresses have been obtained recently
in the BPS protected sector by studying M5 brane compactifications on various space-time backgrounds \cite{Witten:1997sc,Gaiotto:2009we}. In this context, the study of supersymmetric gauge theories via localization and BPS
state counting has revealed to be a very powerful tool \cite{nikitaM}. On one hand this has produced new correspondences among quantum field theories,
topological theories and two-dimensional conformal field theories, as for instance \cite{Alday:2009aq}. On the other it has stimulated the study of higher dimensional supersymmetric gauge theories as deformations
of strongly coupled super-conformal field theories in six dimensions. BPS state counting in this case has been used to capture informations about the circle compactification of M5 branes
in terms of supersymmetric indices \cite{Lockhart:2012vp,Kim:2012qf}. 

In this paper we address the problem of circle compactification of M5 brane systems transverse to an ALE orbifold singularity, which encodes
indices of six-dimensional $\mathcal{N}=(1,0)$ superconformal theories.   
We calculate those indices  for a single M5 via letter counting and show that they coincide with partition functions of suitable five-dimensional quiver gauge theories.
The five-dimensional gauge theories we consider fulfil modular properties which, after T-dualizing to type IIB, reveal to be encoded in the S-duality properties of
a pq-web five-brane system. These latter are crucial in order to expose the enhancement of global symmetries induced by instanton operators. More precisely, we identify a
set of basic five-dimensional instanton operators which generates the full tower of non-perturbative corrections and allows to write the supersymmetric partition functions in a plethystic
exponential form. By going to the S-dual frame we are then able to write these plethystic formulae in terms of characters of the expected enhanced global symmetries.
The main result we obtain is the comparison of the 6d $\mathcal{N}=(1,0)$ superconformal index
for a tensor multiplet and $k^2$ hypermultiplets, computed in Section \ref{SCFI},
with the $S^5$ partition function of a necklace quiver with $k$ abelian nodes \eqref{final-res}. This generalizes the result of Lockart and Vafa \cite{Lockhart:2012vp} for $k=1$.

 One crucial issue in the $5d$ computation is the presence of \emph{spurious} terms, associated to parallel external legs in the pq-web: for a pq-web on a cylinder we show that there are infinite towers of spurious terms. Once we remove these contributions, the $5d$ partition function can be written in terms of $G_2$ special functions \cite{felder1999}. Using the modularity properties, we build the $S^5$ partition function.

These results  have also a nice interpretation in terms of representation theory of $q$-deformed infinite-dimensional Lie algebrae.
Five-dimensional quiver gauge theories partition functions can be interpreted as correlators of vertex operators of $q\mathcal{W}$-algebrae. The pq-web S-duality suggests
relations among correlators of different $q\mathcal{W}$-algebrae, which we check in some examples in Section \ref{qW}.

{\bf Note added:} While this paper was being finalized, \cite{Buican:2016hpb} appeared. Their expressions for the superconformal index of the free $(1,0)$ supermultiplets agree with ours in Section \ref{SCFI}.

\section{M5 branes on $\IC^2 / {\IZ}_k$: 6d and 5d gauge theory descriptions}\label{inst}
We start from $N$ M5 branes sitting at the tip of the orbifold $\IC^2/\IZ_k$ singularity. This is an interacting superconformal $(1,0)$ field theory that we call $\CT^{6d}_{k,A_N}$. 

We can gain some knowledge about this class of theories reducing to Type IIA on a circle inside the $\IC^2/\IZ_k$: the M5's become NS5 branes and the orbifold geometry $\IC^2/\IZ_k$ becomes $k$ D6 branes. So we end up with the following brane setup
\cite{Brunner:1997gk,Brunner:1997gf,Hanany:1997gh}: $N$ NS5 branes sitting on top of a stack of $k$ D6's. 

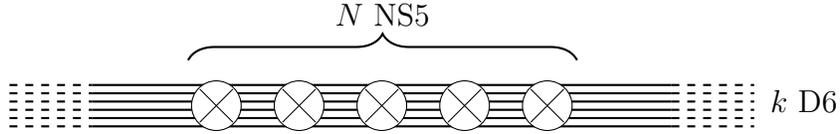
\begin{figure}[h!]
\centering
\begin{tikzpicture}[scale=1]
\foreach \y in {-0.3,-0.2,...,0.3}
{
  \draw[thick]  (-3.5,\y) -- (3.5,\y);
  \draw[thick, dashed]  (3.5,\y) -- (4.5,\y);
  \draw[thick, dashed]  (-4.5,\y) -- (-3.5,\y);
}
\foreach \x in {-2,-1,...,2}
{
\draw[fill=white] (\x,-0.05) circle (0.3);
\draw (\x-0.2,-0.05-0.2) -- (\x+0.2,-0.05+0.2);
\draw (\x-0.2,-0.05+0.2) -- (\x+0.2,-0.05-0.2);
}
\draw[thick] [decorate,decoration={brace,amplitude=10pt},xshift=-4pt,yshift=0pt]
(-2.2,0.5) -- (2.5,0.5) node [black,midway,yshift=0.6cm] 
{$N$ NS5};
\node at (5.1,0) {$k$ D6};
\end{tikzpicture}
\caption{$N$ NS5 branes on top of $k$ D6 branes.}\label{fig1}
\end{figure}

If we separate the NS5 branes we go on the nowadays called tensor branch \cite{DelZotto:2014hpa}, which gives a Lagrangian IR description of the deformed SCFT. On the tensor branch the field theory is a linear quiver $SU(k)^{N-1}$ of the form 

\be
\begin{tikzpicture}[baseline={(A.base)}]
\node[flavor] (A) at (0,0) {$k$};
\node[gauge] (B) at (1,0) {$k$};
\node[gauge] (C) at (2,0) {$k$};
\draw[thick] (A)--(B) ;
\draw[thick] (B)--(C) ;
\draw[thick] (C)--(2.5,0) ;
\end{tikzpicture}\,\cdots\cdots\,
\begin{tikzpicture}[baseline={(A.base)}]
\node[gauge] (B) at (0,0) {$k$};
\node[gauge] (C) at (1,0) {$k$};
\node[flavor] (A) at (2,0) {$k$};
\draw[thick] (B)--(-0.5,0) ;
\draw[thick] (A)--(C) ;
\draw[thick] (B)--(C) ;
\end{tikzpicture}\ .
\ee

There are also $N$ tensor multiplets parameterizing the positions of the $N$ NS5 branes. Both from the Type IIA brane setup and from the quiver, it is easy to see the global symmetry $SU(k)^2$. There is also an additional global $U(1)$ symmetry, that acts on all the $N$ bifundamentals hypers with charge $1$. For $N=1$ the theory is free, we will compute its superconformal index and compare it with $5d$ partition functions. Notice that it is not the usual gauge theory orbifold of the $(2,0)$ free supermultiplet.
 
\subsection*{Compactification to $5d$: S-duality for the rectangular pq-web.}
Reducing $N$ M5 branes sitting at the tip of the orbifold $\IC^2/\IZ_k$ singularity on a circle transverse to $\IC^2/\IZ_k$ and along the M5's, we get $N$ D4 branes sitting at the tip of the orbifold $\IC^2/\IZ_k$ singularity, so the gauge theory can be understood using the methods of \cite{Douglas:1996sw}. One alternative description is given in terms of a pq-web of 5 branes in type IIB: the branes are on a cilinder $\IR \times S^1$, $k$ D5 branes along $\IR$ and $N$ NS5 branes along the $S^1$. If this pq-web was on $\IR^2$ the gauge theory would have been the $SU(N)^{k-1}$ linear quiver, while putting the pq-web on the cilinder the field theory becomes the 5d $\CN=1$ circular quiver $SU(N)^k$ gauge theory, that we call $\CT^{5d}_{k, A_N}$
\be ||
\begin{tikzpicture}[baseline={(A.base)}]
\node[gauge] (B) at (1,0) {$N$};
\node[gauge] (C) at (2,0) {$N$};
\draw[thick] (B)--(C) ;
\draw[thick] (B)--(0.5,0) ;
\draw[thick] (C)--(2.5,0) ;
\end{tikzpicture}\,\cdots\cdots\,
\begin{tikzpicture}[baseline={(A.base)}]
\node[gauge] (B) at (0,0) {$N$};
\node[gauge] (C) at (1,0) {$N$};
\draw[thick] (B)--(-0.5,0) ; 
\draw[thick] (C)--(1.5,0) ; 
\draw[thick] (B)--(C) ;
\end{tikzpicture}
||
\ee

\subsection*{The pq-web on $\IR^2$ and 5d S-duality.}
\label{pq-web-rect}

Let us first analyze the brane setup on $\IR^2$, a pq-web of $N$ D5's intersecting $k$ NS5's
\cite{Hanany:1996ie, Aharony:1997bh}. The gauge theory is a linear quiver $SU(N)^{k-1}$ with $N$ flavors at both ends:
\be
\begin{tikzpicture}[baseline={(A.base)}]
\node[flavor,baseline=(A.base)] (A) at (0,0) {$N$};
\node[gauge] (B) at (1,0) {$N$};
\node[gauge] (C) at (2,0) {$N$};
\draw[thick] (A)--(B) ;
\draw[thick] (B)--(C) ;
\draw[thick] (C)--(2.5,0) ;
\end{tikzpicture}\,\cdots\cdots\,
\begin{tikzpicture}[baseline={(A.base)}]
\node[gauge] (B) at (0,0) {$N$};
\node[gauge] (C) at (1,0) {$N$};
\node[flavor] (A) at (2,0) {$N$};
\draw[thick] (B)--(-0.5,0) ;
\draw[thick] (A)--(C) ;
\draw[thick] (B)--(C) ;
\end{tikzpicture}\ .
\ee

It is also possible to perform a type IIB S-duality on the brane setup, getting the pq-web of $k$ D5's intersecting $N$ NS5's. The gauge theory in this case is the linear quiver $SU(k)^{N-1}$ with $k$ flavors at both ends:
\be
\begin{tikzpicture}[baseline={(A.base)}]
\node[flavor] (A) at (0,0) {$k$};
\node[gauge] (B) at (1,0) {$k$};
\node[gauge] (C) at (2,0) {$k$};
\draw[thick] (A)--(B) ;
\draw[thick] (B)--(C) ;
\draw[thick] (C)--(2.5,0) ;
\end{tikzpicture}\,\cdots\cdots\,
\begin{tikzpicture}[baseline={(A.base)}]
\node[gauge] (B) at (0,0) {$k$};
\node[gauge] (C) at (1,0) {$k$};
\node[flavor] (A) at (2,0) {$k$};
\draw[thick] (B)--(-0.5,0) ;
\draw[thick] (A)--(C) ;
\draw[thick] (B)--(C) ;
\end{tikzpicture}\ .
\ee

The right way to think about this 5d ``duality'' is that there is a strongly coupled 5d SCFT corresponding to the completely unresolved pq-web,
where all branes are on top of each other, emanating from a single point and respecting a rescaling symmetry.
This UV SCFT admits relevant deformations that can lead to either IR Lagrangian QFT: $SU(k)^{N-1}$ or $SU(N)^{k-1}$, which are clearly perturbatively different.
However, if we are able to perform computations in the IR QFT's that can be uplifted to the strongly coupled UV SCFT, like the partition function on $S^4 \times S^1$,
then the results of the two computations should agree \cite{Bao:2011rc, Bergman:2013aca,Mitev:2014jza}.

\subsubsection*{Instanton operators and global symmetry}
5d gauge theories contain non perturbative operators $\I$, charged under the topological symmetries whose currents are $*tr(F^2)$.
When inserted at a point in space-time, the flux of $tr(F^2)$ on the sphere $S^4$ surrounding the point measures the instanton charges of the operators \cite{Yang:1977qv}. 

This is analogous to the 3d case, where monopole operators carry a flux for $tr(F)$ on an $S^2$. In a balanced linear quiver,
there is a special set of `minimal' monopole or instanton operators, see \cite{Benvenuti:2016wet} for a recent discussion in the case of $3d$ $\CN=2$ quivers.
Their topological charges are $0$ or $1$ and are defined by the property that the non vanishing charges are contiguous. For instance for a quiver
$\begin{tikzpicture}
\node[flavor] (A) at (0,0) {$N$};
\node[gauge] (B) at (0.8,0) {$N$};
\node[gauge] (C) at (1.6,0) {$N$};
\node[gauge] (D) at (2.4,0) {$N$};
\node[gauge] (E) at (3.2,0) {$N$};
\node[flavor] (F) at (4,0) {$N$};
\draw[thick] (A)--(B) ;
\draw[thick] (B)--(C) ;
\draw[thick] (C)--(D) ;
\draw[thick] (D)--(E) ;
\draw[thick] (E)--(F) ;
\end{tikzpicture}$
we can organize the $4+3+2+1$ basic instanton states as \be 
\left( \begin{array}{cccc}
\I^{1,0,0,0} & \I^{1,1,0,0} & \I^{1,1,1,0} & \I^{1,1,1,1} \\
               & \I^{0,1,0,0} & \I^{0,1,1,0} & \I^{0,1,1,1} \\
               &             & \I^{0,0,1,0} & \I^{0,0,1,1} \\
               &             &                  & \I^{0,0,0,1} \\
\end{array} \right)
\ee
where the superscripts denote the topological charges of the operator $\I$ under each gauge group.

In 3d for $\CN=4$ theories it is known that these non perturbative operators form a supermultiplet whose primary component has scaling dimension $1$. Such supermultiplets contain conserved currents with scaling dimension $2$. For a quiver $U(N)^{k-1}$, putting together these $k(k-1)/2$ basic monopoles, the correponding $k(k-1)/2$ anti-monopoles and the $k-1$ topological $U(1)$ currents, we get the $k^2-1$ currents of the enhanced $SU(k)$.

In 5d with $\CN=1$ supersymmetry the story should be similar, but it is not much discussed in the literature. Symmetry enhancements of this type have been studied in \cite{Tachikawa:2015mha}, see also \cite{Zafrir:2015uaa,Yonekura:2015ksa,Cremonesi:2015lsa}.

One important feature of these instanton operators is their baryonic charge spectrum: the instanton operators are charged under the Abelian factors of the global symmetries, which are usually called baryonic symmetries. The charges can in principle be computed studying fermionic zero modes around the instanton background. It turns out that a basic instanton state whose topological charges are $1$ from node $i$ to node $j$ is charged precisely under the symmetry that rotates the $i^{th}$ and the $(j+1)^{th}$ bifundamentals. This is the case both in $\CN=2$ 3d theories \cite{Aharony:1997bx, Benvenuti:2016wet} and in $\CN=1$ 5d theories. Let us define $i^{th}$ baryonic symmetry $U(1)_{bar; i}$ to act with charge $+1$ and $-1$ on the $(i-1)^{th}$ and the $i^{th}$ bifundamental, respectively.  Then the basic instantons have baryonic charges equal to $N$ times the topological charges. So, denoting as in \cite{Tachikawa:2015mha},
\be U(1)_{i, \pm} \equiv \frac{1}{2} \left(U(1)_{top; \, i} \pm \frac{U(1)_{bar; i}}{N} \right)\ee
the basic instantons are charged under $U(1)_{i, +} $ and neutral under $U(1)_{i, -}$ . The corresponding anti-instantons are neutral under $U(1)_{i, +} $ and are charged under $U(1)_{i, -} $.

Armed with these results we can study the global symmetries that can be inferred from the low energy Lagrangian description. In the gauge theory $SU(N)^{k-1}$, each gauge group $U(N)$ gives a $U(1)$ ``topological'' or ``instantonic'' global symmetry, whose current is $*tr(F^2)$. Each bifundamental hypermultiplet is charged under a standard $U(1)$ ``baryonic'' symmetry. The theory enjoys a $U(1)^{k-1}_{top} \times U(1)^{k}_{bar} \times SU(N)^2$ global symmetry. This global symmetry is actually enhanced in the UV SCFT. In \cite{Tachikawa:2015mha} it is shown how topological and baryonic symmetries can enhance to a non Abelian group: if we have a IR quiver (or a sub quiver) where every node is balanced, then the global symmetry of the UV SCFT is the square of the group whose Dynkin diagram is the quiver in question. A $U(N)$ node with zero Chern-Simon coupling is balanced if the total number of flavors is precisely $2N$. Here the quiver has the shape of the $A_{k-1}=SU(k)$ Dyinkin diagram so the global symmetry enhancement in the UV is
\be U(1)^{k-1}_{+} \times U(1)^{k-1}_{-}  \rightarrow SU(k)_+ \times SU(k)_- \ee

Starting from the S-dual gauge theory $SU(k)^{N-1}$, we can repeat the same arguments. In both models one concludes that the total global symmetry in the UV is
 \be SU(k)^2 \times SU(N)^2 \times U(1) \ee
This is a well known first check of the pq-web S-duality.

\subsection*{pq-web on the cilinder: 6d/5d duality}
\label{pqweb-cyl}
In the case $k=1$, a well known conjecture \cite{Douglas:2010iu, Lambert:2010iw} relates the 5d $\CN=2$ field theory $\CT^{5d}_{1,A_N}$
to the 6d $(2,0)$ type $A_N$ $\CT^{6d}_{1,A_N}$. 

Here we are adding the orbifold $\IC^2/\IZ_k$, and it is natural to conjecture a relation between $\CT^{5d}_{k,A_N}$ and $\CT^{6d}_{k,A_N}$.

Compactifying the pq-web on a circle we are gauging the $SU(N)$ symmetries together, so the quiver is now the Dynkin diagram of $\widehat{A_k}$, the affine extension of $SU(k)$.

The two $SU(N)$ global symmetries are lost, but we gain one additional topological $U(1)$ 
Also, the sum of all baryonic symmetries acts trivially on the theory. The remaining $U(1)^k_{top} \times U(1)^{k-1}_{bar}$ symmetry is enhanced to the infinite dimensional group $\widehat{A_k\!\!\times\!\!A_k}$, as argued in \cite{Tachikawa:2015mha}. There is also a $U(1)$ symmetry acting on all bifundamentals with charge $+1$.

For the circular quiver there are $k(k-1)$ basic instanton operators with the property that at least one topological charge is zero. We call these `non-wrapping' instantons. There are $k$ `non-wrapping' instantons of length $l$, with $l=1,2,\ldots, k-1$. However thare are also `wrapping' instantons of the form
\be \I^{(1,1,\ldots,1)} \ee
In Section \ref{circular} we will show explicitly how, for $N=1$, wrapping instantons corresponds to Kaluza-Klein modes, summing all of them reproduces a $6d$ $(1,0)$ superconformal index.

\section{6d $(1,0)$ superconformal index}
\label{SCFI}
In this section we derive the superconformal indices for the $6d$ free $(1,0)$ supermultiplets, using letter counting.

The 6d $(1,0)$ and $(2,0)$ superconformal indices are discussed in \cite{Bhattacharya:2008zy}. Here we only need the $(1,0)$ case: the superconformal algebra is $osp(6,2|2)$ with R-symmetry $Sp(2)\simeq SU(2)$. The supercharges $Q^i_\a$ transform in the $(2,4)$ of $SU(2)_R \times SO(6)$.

Picking an appropriate supercharge $Q$ and its conjugate $Q^\dagger$, it is possible to define a Witten index, with fugacities associated to the symmetries that commute with $Q$ and $Q^\dagger$. The index reads
\be \CI = tr (-1)^F q_0^{J_{12}+R}q_1^{J_{34}+R}q_2^{J_{56}+R}  . \ee

Only states with $\{Q,Q^\dagger\}=\delta=0$ contribute to the superconformal index, where 
\be \delta = \Delta - J_{12} - J_{34} - J_{56} - 4 R .  \ee
$\Delta$ is the scaling dimension of the states, $J_{i\, i+1}$ are the angular momenta on the three hortogonal planes in $\IR^6$, $R$ is the $SU(2)_R$ spin.

Let us study explicitly the free superconformal multiplets: hypermultiplet $\{\Phi, \psi\}$ and self-dual tensor multiplet $\{H^{+}, \eta, \phi\}$.

  \begin{center}  \begin{tabular}{|c|c|c|c|c|c|c|c|c|}   \hline
  Letter & $\Delta$ & $J_{12}$ & $J_{34}$ & $J_{56}$  & $R$ & $SO(6)$ irrep & $\delta\!=\!\Delta\!\! -\!\sum \!J\! -\! 4R\!$ & Index \\ \hline\hline
 $\Phi$ & $2$  & $0$ & $0$ & $0$ & $\pm 1/2$ &$1$ & $2 \pm 2$ & $\sqrt{q_0q_1q_2}$ \\ \hline
 $\psi$ & $5/2$  & $1/2$ & $1/2$ & $-1/2$ & $0$ &$4$& $2$ & $0$ \\ \hline
 $\psi$ & $5/2$  & $1/2$ & $-1/2$ & $1/2$ & $0$ &$4$& $2$ & $0$ \\ \hline
 $\psi$ & $5/2$  & $-1/2$ & $1/2$ & $1/2$ & $0$ & $4$& $2$ & $0$ \\ \hline
 $\psi$ & $5/2$  & $-1/2$ & $-1/2$ & $-1/2$ & $0$ &$4$& $4$ & $0$ \\ \hline\hline
 $\phi$ & $2$  & $0$ & $0$ & $0$ & $0$ &$1$ & $2$ & $0$ \\ \hline
 $\eta$ & $5/2$  & $1/2$ & $1/2$ & $-1/2$ & $\pm1/2$ &$4$& $2 \pm 2$ & $-q_0q_1$ \\ \hline
 $\eta$ & $5/2$  & $1/2$ & $-1/2$ & $1/2$ & $\pm1/2$ &$4$& $2 \pm 2$ & $-q_0q_2$ \\ \hline
 $\eta$ & $5/2$  & $-1/2$ & $1/2$ & $1/2$ & $\pm1/2$ &$4$& $2 \pm 2$ & $-q_1q_2$ \\ \hline
 $\eta$ & $5/2$  & $-1/2$ & $-1/2$ & $-1/2$ & $\pm 1/2$ &$4$& $4 \pm 2$ & $0$ \\ \hline
 $H^+$ & $3$  & $1$ & $1$ & $1$ & $0$ &$10$& $0$ & $q_0q_1q_2$ \\ \hline
 $H^+$ & $3$  & $-1$ & $1$ & $1$ & $0$ &$10$& $2$ & $0$ \\ \hline
 $H^+$ & $3$  & $1$ & $-1$ & $1$ & $0$ &$10$& $2$ & $0$ \\ \hline
 $H^+$ & $3$  & $1$ & $1$ & $-1$ & $0$ &$10$& $2$ & $0$ \\ \hline
 $H^+$ & $3$  & $\pm1$ & $0$ & $0$ & $0$ &$10$& $3\pm1$ & $0$ \\ \hline
 $H^+$ & $3$  & $0$ & $\pm1$ & $0$ & $0$ &$10$& $3\pm1$ & $0$ \\ \hline
 $H^+$ & $3$  & $0$ & $0$ & $\pm1$ & $0$ &$10$& $3\pm1$ & $0$ \\ \hline\hline
  $\partial_{1,2}$ & $1$  & $\pm1$ & $0$ & $0$ & $0$ &$6$& $1\pm1$ & $q_0$ \\ \hline
 $\partial_{3,4}$ & $1$  & $0$ & $\pm1$ & $0$ & $0$ &$6$& $1\pm1$ & $q_1$ \\ \hline
 $\partial_{5,6}$ & $1$  & $0$ & $0$ & $\pm1$ & $0$ &$6$& $1\pm1$ & $q_2$ \\ \hline
  \end{tabular}\label{chargesT1} \end{center}

The two free supermultiplets are simple cases in the list of all possible short unitary representations of the $6d$ minimal susy superconformal algebra $osp(6,2|2)$ \cite{Minwalla:1997ka,Dobrev:2002dt}. The full classification is given in terms of the $SO(6)\simeq SU(4)$ Dynkin labels of the superconformal primary of the entire superconformal multiplet.

In the case of the half hypermultiplet the superconformal primary is the $\Delta=2$ complex scalar $\Phi$, transforming in the $2$ of $SU(2)_R$. Acting with the supercharges $Q^i_\a$, we obtain a $SU(2)_R$ singlet fermion $\psi$ with $\Delta=5/2$, transforming in the $4$ of $SO(6)$, while the $SU(2)_R$-triplet is a null state.

In the case of the self-dual tensor multiplet, the superconformal primary is the $\Delta=2$ real scalar $\phi$, an $SU(2)_R$-singlet. Acting with the supercharges $Q^i_\a$, we obtain a $SU(2)_R$-doublet fermion $\eta$ with $\Delta=5/2$, transforming in the $4$ of $SO(6)$. Acting on $\eta$ there is a null state (recall that for $SO(6)$, $4 \otimes 4= 6 \oplus 10$) in the $6$ of $SO(6)$ and the self-dual tensor in the $10$ of $SO(6)$.

Using Table \ref{chargesT1}, the two indices are computed
\be\label{scihyper} \CI_{1/2 hyper} = \frac{\sqrt{q_0q_1q_2}}{(1-q_0)(1-q_1)(1-q_2)} \ee
\be\label{scitensor} \CI_{SD tensor} = \frac{q_0q_1q_2-q_0q_1-q_1q_2-q_0q_2}{(1-q_0)(1-q_1)(1-q_2)} \ee

In Section \ref{circular} it will be easy, using these results, to write down the superconformal index of the $(1,0)$ SCFT corresponding to $1$ M5 brane at the $\IC^2/\IZ_k$ orbifold, that on the tensor branch is simply $k^2$ free hypers plus $1$ self-dual tensor.


\section{Exact partition functions: $5d$ Abelian linear quiver}
\label{linear-quiver}

For $N=1$ it is possible to compute the Nekrasov instanton partition \cite{Nekrasov:2002qd} function explicitly to all orders in the instanton fugacities. 
We review the definition of the Nekrasov partition function in Appendix \ref{nek}. We first consider the simpler case of the linear quiver and we postpone the discussion on the circular quiver, which is the main result of this paper, to the next section. Although the 5d Nekrasov partition function of the linear quiver can be inferred from to the topological string amplitudes computed in \cite{Iqbal:2004ne,Iqbal:2007ii,Taki:2007dh,Bao:2011rc,Mitev:2014jza},
here we perform a direct gauge theory calculation.

 We need to take the multi-particle partition function generated by the basic instantons  we described in Section \ref{inst}.
 This is done by using the so called Plethystic Exponential $PE[f]$ \cite{Benvenuti:2006qr}:
\be
PE[f(t_1, t_2, \ldots, t_K)] = \exp\left[\sum_{n=1}^\infty \frac{f(t_1^n, t_2^n, \ldots, t_K^n)}{n}\right].
\ee

\begin{figure}[h]\label{pqweb-R2}
\centering
\begin{tikzpicture}[scale=0.5,rotate=270]
\foreach \x [count=\xi]  in {0,1,2,3}
{
\draw[thick, dashed] (\x-3,4*\x) -- (\x-2,4*\x); 
\draw[thick] (\x-2,4*\x) -- (\x,4*\x);  
\draw[thick] (\x,4*\x-2) -- (\x,4*\x);  
\draw[thick] (\x,4*\x) node [xshift=17] {$m_{\xi}$} -- (\x+1,4*\x+1);  
\draw[thick] (\x+1,4*\x+1) -- (\x+1,4*\x+3) node [below] {$q_\xi$};   
\draw[thick] (\x+1,4*\x+1) -- (\x+3,4*\x+1);   
\draw[thick, dashed] (\x+3,4*\x+1) -- (\x+4,4*\x+1); 
}
\fill [white] (4.1,14.5) rectangle (4.9,15.5);
\draw[thick, dashed, white] (0,-2) -- (0,-1); 
\draw[thick, dashed, white] (4,14) -- (4,15); 


\draw (9,-1) -- (9,3) -- (9,7) -- (9,11) -- (9,15);
\foreach \x [count=\xi] in {0,1,2,3}
{
\node at (8.5,1+4*\x) {$m_{\xi}$};
}
\foreach \y  in {1,2,3}
{
\node at (7.8,-1+4*\y) {$q_{\y}$};

\draw [thick, fill=white] (9,-1+4*\y) circle (0.8);
\node[font=\large] at (9,-1+4*\y) {$1$};
}

\draw [thick, fill=white] (9-.7,-1-.7) rectangle (9+.7,-1+.7);
\node[font=\large] at (9,-1) {$1$};

\draw [thick, fill=white] (9-.7,15-.7) rectangle (9+.7,15+.7);
\node[font=\large] at (9,15) {$1$};

\end{tikzpicture}
\caption{1 D5 and $k=4$ NS5 on the plane.}
\end{figure}
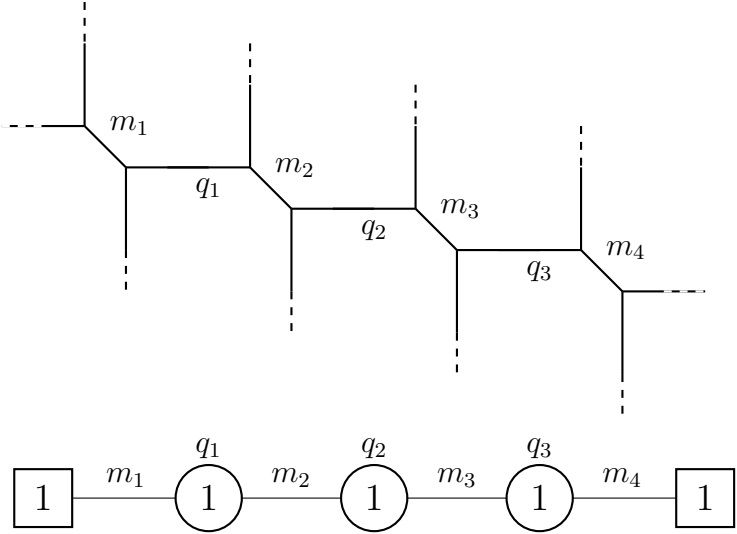

For the Abelian linear quiver with $k-1$ nodes, there are $k(k-1)/2$ basic instantons with topological charges 
$$\I^{(0,0,\ldots,0,1,1,\ldots,1,0,\ldots,0)}.$$
The formula is thus the $PE$ of a sum of $k(k-1)/2$ terms:
\be\label{ZLk}
\Z^{\IR^4 \times S^1}_{linear, inst} = PE \left[ \frac
{\sum_{i=1}^{k-1} \sum_{l=0}^{k-1} (\prod_{s=i}^{i+l}q_s)( \prod_{r=i}^{i+l-1}m_r)(1-m_{i-1}) (1-m_{i+l}t_1t_2)}
{(1-t_1)(1-t_2)}\right]
\ee
$l=0,1,\ldots, k-1$ is the length of the basic instantons, i.e. the number of non zero topological charges.

It is easy to check that this formula correctly reproduces the terms proportional to one single $q_i$ and to $q_iq_{i+1}$ in the Nekrasov partition function.
\eqref{ZLk} can also be checked to high orders in the instanton fugacities $q_i$ with Mathematica against the Nekrasov partition function.
\eqref{ZLk} must be supplemented by the perturbative contribution of $k$ hypers:
\be
\Z^{\IR^4 \times S^1}_{linear, pert} 
= PE \left[ \frac{t_1t_2\sum_{i=1}^k m_i }{(1-t_1)(1-t_2)}\right]
\ee

\begin{figure}
\centering
\begin{tikzpicture}[scale=0.6]

\draw[dashed,very thick] (0,6) -- (1,6);   
\draw[very thick] (1,6) -- 
(3,6) node[right] {$m_1$} -- (4,5) node[xshift=20pt,yshift=5pt] {$q_1$} --
(6,5) node[right] {$m_2$} -- (7,4) node[xshift=20pt,yshift=5pt] {$q_2$} --
(9,4) node[right] {$m_3$} -- (10,3) -- (12,3);
\draw[dashed,very thick] (12,3) -- (13,3); 

\draw[very thick] (3,6) -- (3,9.5) (6,5) -- (6,9) (9,4) -- (9,8.5); 
\draw[very thick,dashed] (3,9.5) -- (3,10.5) (6,9) -- (6,10) (9,8.5) -- (9,9.5); 
\draw[very thick] (4,5) -- (4,0.5) (7,4) -- (7,0) (10,3) -- (10,-0.5); 
\draw[very thick,dashed] (4,0.5) -- (4,-0.5) (7,0) -- (7,-1) (10,-0.5) -- (10,-1.5); 


\draw[thick,red] (3,7) -- (4.5,7) node[above] {$m_1 q_1$} -- (6,7);
\draw[thick,red] (6,6.5) -- (7.5,6.5) node[above] {$m_2 q_2$} -- (9,6.5);
\draw[thick,red] (3,8) -- (4.5,8) node[above] {$m_1 m_2 q_1 q_2$} -- (9,8);

\fill [fill=white] (6,8) circle (0.2);   
\draw[very thick] (6,8-0.2) -- (6,8+0.2);
\draw[thick,red] (6+0.2,8) arc (0:180:0.2);


\draw[thick,blue] (4,3) -- (5.5,3) node[above] {$m_2 q_1$} -- (7,3);
\draw[thick,blue] (7,2.5) -- (8.5,2.5) node[above] {$m_3 q_2$} -- (10,2.5);
\draw[thick,blue] (4,1) -- (5.5,1) node[above] {$m_2 m_3 q_1 q_2$} -- (10,1);

\fill [fill=white] (7,1) circle (0.2);   
\draw[very thick] (7,1-0.2) -- (7,1+0.2);
\draw[thick,blue] (7+0.2,1) arc (0:180:0.2);

\end{tikzpicture}
\caption{Graphical representation of the $k(k-1)$ spurious contributions. These are D1 branes that can slide off to infinity. In the picture $k=3$.}
\label{slideoff}
\end{figure}
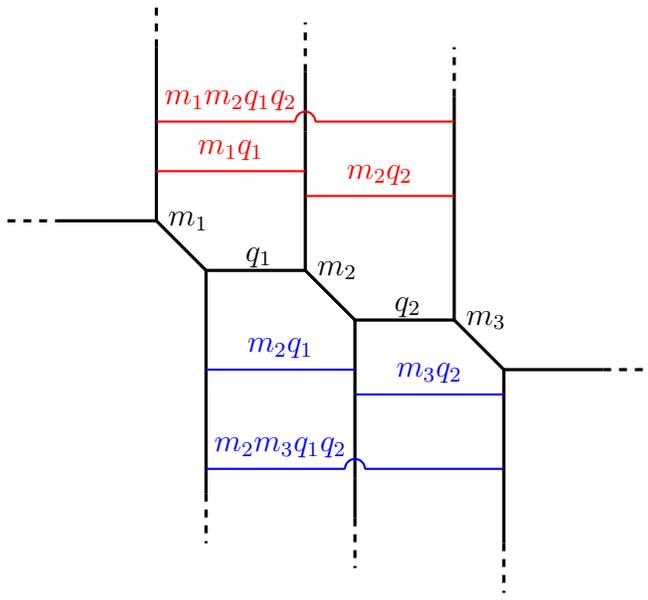

We get, for $\Z^{\IR^4 \times S^1}_{linear, pert+inst}$ 
\be PE \left[ \frac
{(\sum_{i=1}^k \tm_i) t_1t_2+ \sum_{i,I=1}^{k-1}(\prod_{s=i}^{i+I-1}\qt_s\mt_s) (1-\tm_{i-1}) (\tm_{i+I-1}^{-1}-t_1t_2)}
{(1-t_1)(1-t_2)}\right] \ee
Each of the $k(k-1)/2$ instantonic terms decomposes in two positive terms and two negatives terms, so in the double sum in the numerator we can collect
\begin{itemize} 
\item $k^2$ positive terms, $k(k-1)$ instanton plus $k$ perturbative terms. We will show shortly that the positive terms transform like the bifundamental of the enhanced global symmetries $SU(k) \times SU(k)$:
\be
\sum_{i=1}^k \tm_it_1t_2 + \sum_{i,I=1}^{k-1}(\prod_{s=i}^{i+I-1}\qt_s\mt_s) (\tm_{i-1}t_1t_2+\tm_{i+I-1}^{-1}) \ee
\item $k(k-1)/2$ negative terms of the form $-\qt_i \tm_{i-1}, -\qt_i \tm_{i-1} \qt_{i+1} \tm_{i}, \ldots $. These terms transform like `half' adjoints (positive roots) of one of the two $SU(k)$ factors.
\item $k(k-1)/2$ negative terms of the form $-\qt_i \tm_{i} t_1t_2, -\qt_i \tm_{i} \qt_{i+1} \tm_{i+1}t_1t_2, \ldots $.
These terms transform like `half' adjoints (positive roots) of the other $SU(k)$ factor.
\end{itemize} 
The latter $k(k-1)$ negative terms, beside being negative, are not invariant under $t_i \rightarrow 1/t_i$, they must be removed. We call these negative terms \emph{spurious}.
The spurious contributions come from D1 branes stretching between parallel external NS5's in the pq-web, that can slide off to infinity,
as was understood in \cite{Bergman:2013ala,Bao:2013pwa,Hayashi:2013qwa,Bergman:2013aca,Mitev:2014jza}, see Fig.~\ref{slideoff}.

We now want to show that \eqref{ZLk}, after removing the spurious contributions, reproduces precisely the partition function of $k^2$ free hypers, a expected from S-duality.
We construct the $S^4 \times S^1$ partition functions multiplying the contribution from the North and South poles, that are the $\IR^4 \times S^1$ Nekrasov partition functions.
We analyze the cases $k=1$ (which is trivial) and $k=2$ first.

\subsection*{$k=1$} 
In this case we just have the perturbative contribution of a free hyper
\be
\Z^{\IR^4 \times S^1}_{k=1, full}=\Z^{\IR^4 \times S^1}_{k=1, pert}  = PE \left[ \frac{ \tilde m_0\sqrt{t_1t_2} }{(1-t_1)(1-t_2)}\right] . 
\ee
where $\tilde m_0=m_0\sqrt{t_1t_2}$. On $S^4 \times S^1$ we get
\be
\Z^{S^4 \times S^1}_{k=1, full} =PE \left[ 
\frac{\tilde  m_0\sqrt{t_1t_2} }{(1-t_1)(1-t_2)} + \frac
{\tilde  m_0^{-1}\sqrt{t_1^{-1}t_2^{-1}} }{(1-t_1^{-1})(1-t_2^{-1})}\right] 
= PE \left[\frac{ (\tilde  \tm_0 + \tilde  \tm_0^{-1})\sqrt{t_1t_2}}{(1-t_1)(1-t_2)} \right]
\ee
This is the $S^4 \times S^1$ 5d index of a free hyper with mass $\tilde  \tm_0$.

\subsection*{$k=2$}
In this case the linear quiver is 
$\begin{tikzpicture}
\node[flavor] (A) at (0,0) {$1$};
\node[gauge] (B) at (0.8,0) {$1$};
\node[flavor] (C) at (1.6,0) {$1$};
\draw[thick] (A)--(B) ;
\draw[thick] (B)--(C) ;
\node at (2.1,-.1) {$=$};
\node[gauge] (D) at (2.6,0) {$1$};
\node[flavor] (E) at (3.4,0) {$2$};
\draw[thick] (D)--(E) ;
\end{tikzpicture}$
\be
\Z^{\IR^4 \times S^1}_{k=2, pert+inst} 
= PE \left[ \frac
{ (\tm_0 + \tm_1)t_1t_2 + \qt  (1  - \tm_{0})(1 -\tm_{1}t_1t_2 )}
{(1-t_1)(1-t_2)}\right]\\
\ee
Where $\Z^{\IR^4 \times S^1}_{k=2, pert+inst}=\Z^{\IR^4 \times S^1}_{k=2, pert}\Z^{\IR^4 \times S^1}_{k=2, inst}$.
Changing variables to 
\be
\tm_0 = \frac{xA}{y\sqrt{t_1t_2}} \qquad \tm_1=\frac{yA}{x\sqrt{t_1t_2}} \qquad 
\qt=\frac{xy\sqrt{t_1t_2}}{A}
\ee
with inverse
\be
x= \sqrt{\tm_0\qt} \qquad  A= \sqrt{\tm_0\tm_1 t_1 t_2} \qquad y= \sqrt{\qt\tm_1}
\ee
we get
\be
\Z^{\IR^4 \times S^1}_{k=2, pert+inst} = PE \left[ \frac
{ (xA/y + Ay/x + xy/A + xyA )\sqrt{t_1t_2} - x^2 - y^2 t_1t_2}
{(1-t_1)(1-t_2)}\right]
\ee
On $S^4 \times S^1$ we get
\be
\Z^{S^4 \times S^1}_{k=2, pert+inst} = PE \left[ \frac
{ (x + 1/x )( A+1/A)(y+1/y)\sqrt{t_1t_2} - (x^2 + 1/y^2 + (1/x^2 + y^2 )t_1t_2) }
{(1-t_1)(1-t_2)}\right]
\ee
In the 8 positive terms we recognise the trifundamental of $SU(2)^3$, with fugacities $x,A,y$.
The 4 negative terms are spurious and must be removed.
Recall that in the case $k=2$
$\begin{tikzpicture}
\node[flavor] (B) at (1,0) {$2$};  \node[flavor] (C) at (2,0) {$2$};
\draw[thick] (B)--(C);
\end{tikzpicture}$, with $SU(2)^2 \times U(1)$ global symmetry,
is actually a trifundamental
$\begin{tikzpicture}
\node[flavor] (A) at (-0.4,-0.3) {$2$};  
\node[flavor] (B) at (0.4,-0.3) {$2$};  
\node[flavor] (C) at (0,0.5) {$2$};
\draw[thick] (A)--(0,0);
\draw[thick] (B)--(0,0);
\draw[thick] (C)--(0,0);
\end{tikzpicture}$. So we recover the partition function of $4$ free hypers, as expected from S-duality.

Removing the spurious negative terms on $\IR^4 \times S^1$ amounts to multiply the partition function by a factor
\be
\Z^{\IR^4 \times S^1}_{k=2, spurious} = PE \left[ \frac{ \qt \mt_0 + \qt \mt_1 t_1t_2}{(1-t_1)(1-t_2)}\right].
\ee

\subsection*{Generic $k$}
Let us change variables in \eqref{ZLk} from the $k$ masses $m_i$ and $k-1$ couplings $q_i$ to $x_i$, $y_i$ ($i=1,\ldots,k-1$, $x_0=y_0=x_k=y_k=1$) and $A$:
\be  
m_i = \frac{x_{i+1}y_{i}A}{x_{i}y_{i+1}\sqrt{t_1t_2}} \qquad  
q_i = \frac{x_i y_i\sqrt{t_1t_2}}{x_{i+1}y_{i-1}A}
\ee
which implies
\be
\begin{aligned}
\prod_{s=j}^{j+I-1}q_s m_s &= \frac{y_{j}}{y_{j-1}}\frac{y_{j+I-1}}{y_{j+I}} &\qquad 
\prod_{s=j}^{j+I-1}q_s m_{s-1} &= \frac{x_{j}}{x_{j-1}}\frac{x_{j+I-1}}{x_{j+I}}
\end{aligned}
\ee 
The $x_i$ and $y_i$ will be the chemical potentials of the enhanced $SU(k) \times SU(k)$ symmetry, and $A$ is the chemical potential of the $U(1)$ symmetry. Recalling that our formula for generic $k$ is
\be PE \left[ \frac
{(\sum_{i=1}^k \tm_i) t_1t_2+ \sum_{i,I=1}^{k-1}(\prod_{s=i}^{i+I-1}\qt_s\mt_s) (1-\tm_{i-1}) (\tm_{i+I-1}^{-1}-t_1t_2)}
{(1-t_1)(1-t_2)}\right] \ee
under the above change of variables, the positive terms in the numerator become
\be 
\sum_{j=1}^k \frac{x_{j+1}y_{j}}{x_{j}y_{j+1}}A\sqrt{t_1t_2}  +\sum_{j=1}^{k-1}\sum_{I=1}^{k-j}
 \Big(\frac{x_j y_{j+I-1}}{x_{j-1}y_{j+I}}A + \frac{y_j x_{j+I-1}}{y_{j-1}x_{j+I}}A^{-1} \Big) \sqrt{t_1t_2} \,,
 \ee
while the negative terms become
\be 
 \sum_{j=1}^{k-1}\sum_{I=1}^{k-j}\prod_{s=j}^{j+I-1}\qt_s\mt_s (\mt_{j-1}\mt_{j+I-1}^{-1}+t_1t_2 )= 
\sum_{j=1}^{k-1}\sum_{I=1}^{k-j}
 \Big(\frac{x_{j}}{x_{j-1}}\frac{x_{j+I-1}}{x_{j+I}}+\frac{y_{j}}{y_{j-1}}\frac{y_{j+I-1}}{y_{j+I}}t_1t_2 \Big) \, . 
 \ee

The negative terms are associated to strings stretching between external D5 branes, so they must be removed.
Their flavour fugacities are precisely the ones expected from the pq-web, and transform like a `half-adjoint' of the two $SU(k)$ groups, while they are not charged under the $U(1)$.

We are then left with $k^2$ positive terms, $k(k+1)/2$ terms with $A$-charge `$+1$', $k(k-1)/2$ terms with $A$-charge `$-1$'. 

When we consider the gauge theory on $S^4 \times S^1$ we need to sum, inside the PE, the contribution from the North pole and the South pole.
The South pole contribution has all the chemical potentials, and $t_1$, $t_2$, inverted, so we get $k^2$ terms with $A$-charge `$+1$', $k^2$ terms with $A$-charge `$-1$'.
The full $S^4 \times S^1$ partition function $\Z_{full}=\Z_{pert}\Z_{inst}\Z_{spurious}$ can be simplified to
\ben\label{linearfinal}
\Z^{S^4 \times S^1}_{linear,full} &=& PE\left[ \frac
{((\sum x_{i+1}/x_i)(\sum y_j/y_{j+1})A + (\sum x_i/x_{i+1})(\sum y_{j+1}/y_j)A^{-1})\sqrt{t_1t_2} }{(1-t_1)(1-t_2)}\right] \nn \\
&=& PE \left[ \frac
{ (\chi^{SU(k)}_{fund}[x_i]\chi^{SU(k)}_{antifund}[y_i]  A+ \chi^{SU(k)}_{antifund}[x_i]\chi^{SU(k)}_{fund}[y_i] A^{-1}) \sqrt{t_1t_2} }
{(1-t_1)(1-t_2)}\right]
\een
where $\chi^{SU(k)}_{(anti)fund}[x_i]$ is the character of the (anti)fundamental representation of $SU(k)$.
Eq. \eqref{linearfinal} is the same partition function of $k^2$ free hypers in the bifundamental 
$\begin{tikzpicture}
\node[flavor] (B) at (0,-0.5) {$k$};  \node[flavor] (C) at (0.9,-0.5) {$k$};
\draw[thick] (B)--(C);
\end{tikzpicture}$. 
Summarizing, we proved that
\be
\Z_{[1]\!-\!(1)_{\phantom{a}}^{\!k\!-\!1}\!-\![1]}(q_i,m_j) = \Z_{[k]\!-\![k]}(x_i,y_j,A)
\ee
where the two sets of variables are related by
\be  
m_i = \frac{x_{i+1}y_{i}A}{x_{i}y_{i+1}\sqrt{t_1t_2}} \qquad  
q_i = \frac{x_i y_i\sqrt{t_1t_2}}{x_{i+1}y_{i-1}A}
\ee


\section{Exact partition functions: $5d$ Abelian circular quiver}
\label{circular}
In this section we write down the exact Nekrasov instanton partition function
for the abelian necklace quiver, we study the spurious terms,
we construct the $S^5$ partition function and we
compare it with the 6d $(1,0)$ index.

The topological string partition for this pq-web has been studied in
\cite{Haghighat:2013tka, Shabbir:2015oxa}. 

For the abelian necklace quiver (figures \ref{fig-neck-1} and \ref{fig-neck-2}) the partition function (see appendix \ref{nek}) receives contribution from $k(k-1)$ non wrapping instantons, but we also need to consider the wrapping instantons of the form $\I^{(1,1,\ldots,1)}$. It turns out that, in order to reproduce the Nekrasov partition function, we need to sum over a full tower of wrapping instantons $\I^{(n,n,\ldots,n)}$ for all $n \geq 1$ \footnote{Notice that this aspect looks different from the $3d$ $\CN=2$ case. In \cite{Benvenuti:2016wet} it is shown how in circular quivers with flavors at each node there is only one wrapping monopole in the chiral ring, with all topological charges $+1$. The wrapping monopoles with all charges $+n$ are simply the $n^{th}$ power of the basic one.}.
From the index computation point of view the instanton quantum number 
corresponds to the Kaluza-Klein charge on the circle \cite{Kim:2011mv}.
We are thus led to propose the following formula:
\be \Z^{\IR^4 \times S^1}_{inst} = PE \left[ \frac{Q}{1- Q}+\frac
{\sum_{i,I=1}^k  (\prod_{s=i}^{i+I-1}\qt_s \tm_s) (1-\tm_{i-1}) (\tm_{i+I-1}^{-1}- t_1t_2)}
{(1- Q)(1-t_1)(1-t_2)}\right] \ee
where $Q= \prod_{i=1}^kq_i m_i$ is the lenght of the circle the pq-web lives on. We checked this result to high orders with Mathematica.

There is also the perturbative contribution of the $k$ hypermultiplets
\be
\Z^{\IR^4 \times S^1}_{pert} = PE \left[ \frac{\sum_{i=1}^k \mt_i t_1t_2 }{(1-t_1)(1-t_2)}\right],
\ee


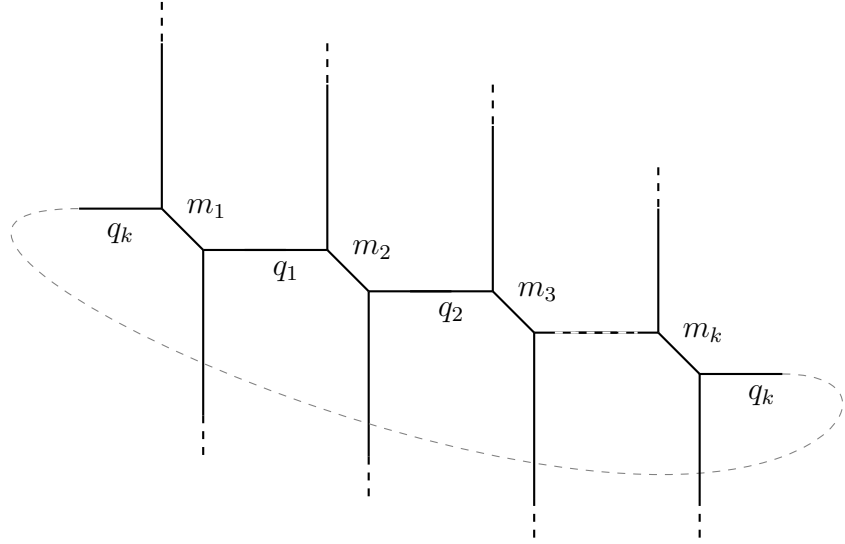
\begin{figure}[h!]
\centering
\begin{tikzpicture}[scale=0.5,rotate=270]

\foreach \x [count=\xi]  in {0,1,2}
{
\draw[thick, dashed] (\x-5,4*\x) -- (\x-4,4*\x); 
\draw[thick] (\x-4,4*\x) -- (\x,4*\x);  
\draw[thick] (\x,4*\x-2) -- (\x,4*\x);  
\draw[thick] (\x,4*\x) node [xshift=17] {$m_\xi$} -- (\x+1,4*\x+1);  
\draw[thick] (\x+1,4*\x+1) -- (\x+1,4*\x+3) node [below] {$q_\xi$};   
\draw[thick] (\x+1,4*\x+1) -- (\x+5,4*\x+1);   
\draw[thick, dashed] (\x+5,4*\x+1) -- (\x+6,4*\x+1); 
}
\fill [white] (3.1,10.5) rectangle (3.9,11.5);

\draw[thin, black!50, dashed] (4,15) to [out=90, in=110, distance=2cm] (6,15) to [out=-70, in=-90] (0,-2);

\foreach \x  in {3}
{
\draw[thick, dashed] (\x-4,4*\x) -- (\x-3,4*\x); 
\draw[thick] (\x-3,4*\x) -- (\x,4*\x);  
\draw[thick] (\x,4*\x-2) -- (\x,4*\x);  
\draw[thick] (\x,4*\x) node [xshift=17] {$m_k$} -- (\x+1,4*\x+1);  
\draw[thick] (\x+1,4*\x+1) -- (\x+1,4*\x+2.5) node [below] {$q_k$} -- (\x+1,4*\x+3);   
\draw[thick] (\x+1,4*\x+1) -- (\x+4,4*\x+1);   
\draw[thick, dashed] (\x+4,4*\x+1) -- (\x+5,4*\x+1); 
}
\draw[thick, dashed, white] (3,9.5) -- (3,11.5); 
\node at (0.6,-1) {$q_k$};
\node[white] at (0,20) {a};
\end{tikzpicture}
\caption{1 D5 and $k$ NS5 on the cylinder}\label{fig-neck-1}
\end{figure}

\begin{figure}[h!]
\centering
\begin{tikzpicture}[scale=0.9]
\draw[thick, dashed] (0,0) circle (3);
\draw[thick] (0,-3) arc (-90:135:3);

\foreach \angle [count=\xi] in {90,45,...,-90}
{
  \node[font=\large] at (\angle:3.8cm) {\textsf{$q_\xi$}};

 \draw [thick, fill=white] (\angle:3) circle (0.5);
 \node[font=\large] at (\angle:3) {$1$};
}
\node[font=\large] at (135:3.8cm) {\textsf{$q_k$}};

 \draw [thick, fill=white] (135:3) circle (0.5);
 \node[font=\large] at (135:3) {$1$};

\foreach \angle [count=\xi] in {90-22.5,45-22.5,-22.5,-45-22.5}
{
  \node[font=\large] at (\angle:3.4cm) {\textsf{$m_\xi$}};
}
\node[font=\large] at (135-22.5:3.4cm) {\textsf{$m_k$}};

\end{tikzpicture}
\caption{Circular quiver with $k$ $U(1)$ nodes.}\label{fig-neck-2}
\end{figure}
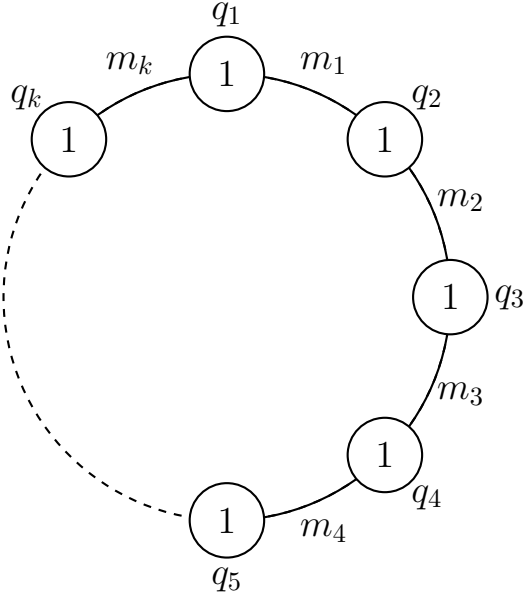

Let us focus on the numerator inside the PE and split the instanton sum into wrapping instantons and non wrapping instantons:
\be \begin{aligned}
(1-Q)(\sum_{i=1}^k \mt_i t_1t_2) + Q(1-t_1)(1-t_2) + Q \sum_{i=1}^k  (1-\tm_{i}) (\tm_{i}^{-1}-t_1t_2) +\\
+\nn \sum_{i=1}^k \sum_{I=1}^{k-1}  (\prod_{s=i}^{i+I-1}\qt_s \tm_s) (1-\tm_{i-1}) (\tm_{i+I-1}^{-1}- t_1t_2) \end{aligned} \ee
Which can be rewritten as
\ben \nn Q(1-t_1)(1-t_2) -k Q t_1t_2 - kQ +\sum_{i=1}^k( \mt_i t_1t_2+ \tm_{i}^{-1}Q)  +\\
+ \sum_{i=1}^k \sum_{I=1}^{k-1}  (\prod_{s=i}^{i+I-1}\qt_s \tm_s) (\tm_{i+I-1}^{-1}+\tm_{i-1}t_1t_2 - t_1t_2 -\tm_{i+I-1}^{-1}\tm_{i-1}) \label{wrap+nonwrap}
\een

\begin{figure}
\centering
\begin{tikzpicture}[scale=0.6]

\draw[dashed,very thick] (1,-1) -- (1,15);  
\draw[very thick] (1-.2,3-.2) -- (1+.2,3) (1-.2,3) -- (1+.2,3+.2);  
\draw[dashed,very thick] (9,-1) -- (9,15);  
\draw[very thick] (9-.2,3-.2) -- (9+.2,3) (9-.2,3) -- (9+.2,3+.2);  

\draw[very thick] (1,6) -- 
(3,6) node[right] {$m_1$} -- (4,5) node[xshift=20pt,yshift=5pt] {$q_1$} --
(6,5) node[right] {$m_2$} -- (7,4) node[xshift=20pt,yshift=5pt] {$q_2$}
 -- (9,4);  
\node[above] at (2,6) {$q_2$};

\draw[very thick] (3,6) -- (3,14) (6,5) -- (6,14); 
\draw[very thick,dashed] (3,14) -- (3,15) (6,14) -- (6,15); 
\draw[very thick] (4,5) -- (4,0) (7,4) -- (7,0); 
\draw[very thick,dashed] (4,0) -- (4,-1) (7,0) -- (7,-1); 


\draw[thick,red] (3,7) -- (4.5,7) node[above] {$m_1 q_1$} -- (6,7);

\draw[thick,red] (6,7.8) -- (7.5,7.8) node[above] {$m_2 q_2$} -- (9,7.8);
\draw[thick,red] (1,7.8) -- (3,7.8);

\draw[thick,red] (3,9) -- (9,9);
\draw[thick,red] (1,9.3) -- (4.5,9.3) node[above] {$m_1 q_1 Q$} -- (6,9.3);

\fill [fill=white] (6,9) circle (0.2);   
\draw[thick,red] (6+0.2,9) arc (0:180:0.2);
\draw[very thick] (6,9-0.2) -- (6,9+0.2);
\fill [fill=white] (3,9.3) circle (0.2);   
\draw[thick,red] (3+0.2,9.3) arc (0:180:0.2);
\draw[very thick] (3,9.3-0.2) -- (3,9.3+0.2);

\draw[thick,red] (6,10.8) -- (9,10.8);
\draw[thick,red] (1,11) -- (7.5,11)   node[above] {$m_2 q_2 Q$}  -- (9,11);
\draw[thick,red] (1,11.3) -- (3,11.3);

\fill [fill=white] (6,11) circle (0.2);   
\draw[thick,red] (6+0.2,11) arc (0:180:0.2);
\draw[very thick] (6,11-0.2) -- (6,11+0.2);
\fill [fill=white] (3,11) circle (0.2);   
\draw[thick,red] (3+0.2,11) arc (0:180:0.2);
\draw[very thick] (3,11-0.2) -- (3,11+0.2);

\draw[thick,red] (3,12.5) -- (9,12.5);
\draw[thick,red] (1,12.5+.2) -- (9,12.5+.2);
\draw[thick,red] (1,12.5+.5) -- (4.5,12.5+.5) node[above] {$m_1 q_1 Q^2$} -- (6,12.5+.5);

\fill [fill=white] (6,12.5) circle (0.2);   
\fill [fill=white] (3,12.5+.2) circle (0.2);
\fill [fill=white] (6,12.5+.2) circle (0.2);
\fill [fill=white] (3,12.5+.5) circle (0.2); 
\draw[thick,red] (6+0.2,12.5) arc (0:180:0.2); 
\draw[very thick] (6,12.5-0.2) -- (6,12.5+0.2);
\draw[thick,red] (3+0.2,12.5+.2) arc (0:180:0.2); 
\draw[very thick] (3,12.5+.2-0.2) -- (3,12.5+.2+0.2); 
\draw[thick,red] (6+0.2,12.5+.2) arc (0:180:0.2); 
\draw[very thick] (6,12.5+.2-0.2) -- (6,12.5+.2+0.2);
\draw[thick,red] (3+0.2,12.5+.5) arc (0:180:0.2); 
\draw[very thick] (3,12.5+.5-0.2) -- (3,12.5+.5+0.2);

\foreach \y in {1,2,3}
\fill[red] (4.5,14+0.3*\y) circle (0.05);


\draw[thick,blue] (4,3) -- (5.5,3) node[above] {$m_2 q_1$} -- (7,3);

\draw[thick,blue] (7,2) -- (9,2);
\draw[thick,blue] (1,2) -- (2.5,2) node[above] {$m_1 q_2$} -- (4,2);

\draw[thick,blue] (4,0.2) -- (9,0.2);
\draw[thick,blue] (1,0.5) -- (5.5,0.5) node[above] {$m_2 q_1 Q$} -- (7,0.5);

\fill [fill=white] (7,0.2) circle (0.2);   
\draw[very thick] (7,0.2-0.2) -- (7,0.2+0.2);
\draw[thick,blue] (7+0.2,0.2) arc (0:180:0.2);
\fill [fill=white] (4,0.5) circle (0.2);   
\draw[very thick] (4,0.5-0.2) -- (4,0.5+0.2);
\draw[thick,blue] (4+0.2,0.5) arc (0:180:0.2);

\foreach \y in {1,2,3}
\fill[blue] (5.5,0-0.3*\y) circle (0.05);


\draw[dashed,very thick] (13,-1) -- (13,15);  
\draw[very thick] (13-.2,3-.2) -- (13+.2,3) (13-.2,3) -- (13+.2,3+.2);  
\draw[dashed,very thick] (21,-1) -- (21,15);  
\draw[very thick] (21-.2,3-.2) -- (21+.2,3) (21-.2,3) -- (21+.2,3+.2);  

\draw[very thick] (13,6) -- 
(15,6) node[right] {$m_1$} -- (16,5) node[xshift=20pt,yshift=5pt] {$q_1$} --
(18,5) node[right] {$m_2$} -- (19,4) node[xshift=20pt,yshift=5pt] {$q_2$}
 -- (21,4);  
\node[above] at (14,6) {$q_2$};

\draw[very thick] (15,6) -- (15,14) (18,5) -- (18,14); 
\draw[very thick,dashed] (15,14) -- (15,15) (18,14) -- (18,15); 
\draw[very thick] (16,5) -- (16,0) (19,4) -- (19,0); 
\draw[very thick,dashed] (16,0) -- (16,-1) (19,0) -- (19,-1); 


\draw[thick,red] (13,7.5) -- (16.5,7.5) node[above] {$Q$} -- (21,7.5);
\fill [fill=white] (18,7.5) circle (0.2);   
\draw[thick,red] (18+0.2,7.5) arc (0:180:0.2);
\draw[very thick] (18,7.5-0.2) -- (18,7.5+0.2);

\draw[thick,red] (13,9) -- (16.5,9) node[above] {$Q$} -- (21,9);
\fill [fill=white] (15,9) circle (0.2);   
\draw[thick,red] (15+0.2,9) arc (0:180:0.2);
\draw[very thick] (15,9-0.2) -- (15,9+0.2);

\draw[thick,red] (15,10.5) -- (21,10.5);
\draw[thick,red] (13,10.5+.2) -- (16.5,10.5+.2) node[above] {$Q^2$} -- (21,10.5+.2);
\draw[thick,red] (13,10.5+.5)  -- (15,10.5+.5);

\fill [fill=white] (18,10.5) circle (0.2);   
\fill [fill=white] (15,10.5+.2) circle (0.2);
\fill [fill=white] (18,10.5+.2) circle (0.2); 
\draw[thick,red] (18+0.2,10.5) arc (0:180:0.2); 
\draw[very thick] (18,10.5-0.2) -- (18,10.5+0.2);
\draw[thick,red] (15+0.2,10.5+.2) arc (0:180:0.2); 
\draw[very thick] (15,10.5+.2-0.2) -- (15,10.5+.2+0.2); 
\draw[thick,red] (18+0.2,10.5+.2) arc (0:180:0.2); 
\draw[very thick] (18,10.5+.2-0.2) -- (18,10.5+.2+0.2);

\draw[thick,red] (18,12.5) -- (21,12.5);
\draw[thick,red] (13,12.5+.2) -- (21,12.5+.2);
\draw[thick,red] (13,12.5+.5) -- (16.5,12.5+.5) node[above] {$Q^2$} -- (18,12.5+.5);
  
\fill [fill=white] (15,12.5+.2) circle (0.2); 
\fill [fill=white] (18,12.5+.2) circle (0.2);
\fill [fill=white] (15,12.5+.5) circle (0.2); 
\draw[thick,red] (15+0.2,12.5+.2) arc (0:180:0.2); 
\draw[very thick] (15,12.5+.2-0.2) -- (15,12.5+.2+0.2); 
\draw[thick,red] (18+0.2,12.5+.2) arc (0:180:0.2); 
\draw[very thick] (18,12.5+.2-0.2) -- (18,12.5+.2+0.2);
\draw[thick,red] (15+0.2,12.5+.5) arc (0:180:0.2); 
\draw[very thick] (15,12.5+.5-0.2) -- (15,12.5+.5+0.2);

\foreach \y in {1,2,3}
\fill[red] (16.5,14+0.3*\y) circle (0.05);


\draw[thick,blue] (13,2.5) -- (17.5,2.5) node[above] {$Q$} -- (21,2.5);
\fill [fill=white] (19,2.5) circle (0.2);   
\draw[very thick] (19,2.5-0.2) -- (19,2.5+0.2);
\draw[thick,blue] (19+0.2,2.5) arc (0:180:0.2);

\draw[thick,blue] (13,1) -- (17.5,1) node[above] {$Q$} -- (21,1);
\fill [fill=white] (16,1) circle (0.2);   
\draw[very thick] (16,1-0.2) -- (16,1+0.2);
\draw[thick,blue] (16+0.2,1) arc (0:180:0.2);

\foreach \y in {1,2,3}
\fill[blue] (17.5,0-0.3*\y) circle (0.05);

\end{tikzpicture}

\caption{Spurious contributions on the cylinder. Here we display $k=2$. On the left the terms dependent on $q_i$ and $m_i$, associated to non-wrapping instantons, second line of \eqref{wrap+nonwrap}. On the right the terms dependent only on $Q$, associated to wrapping instantons, first line of \eqref{wrap+nonwrap}.}\label{spurious-cyl}
\end{figure}
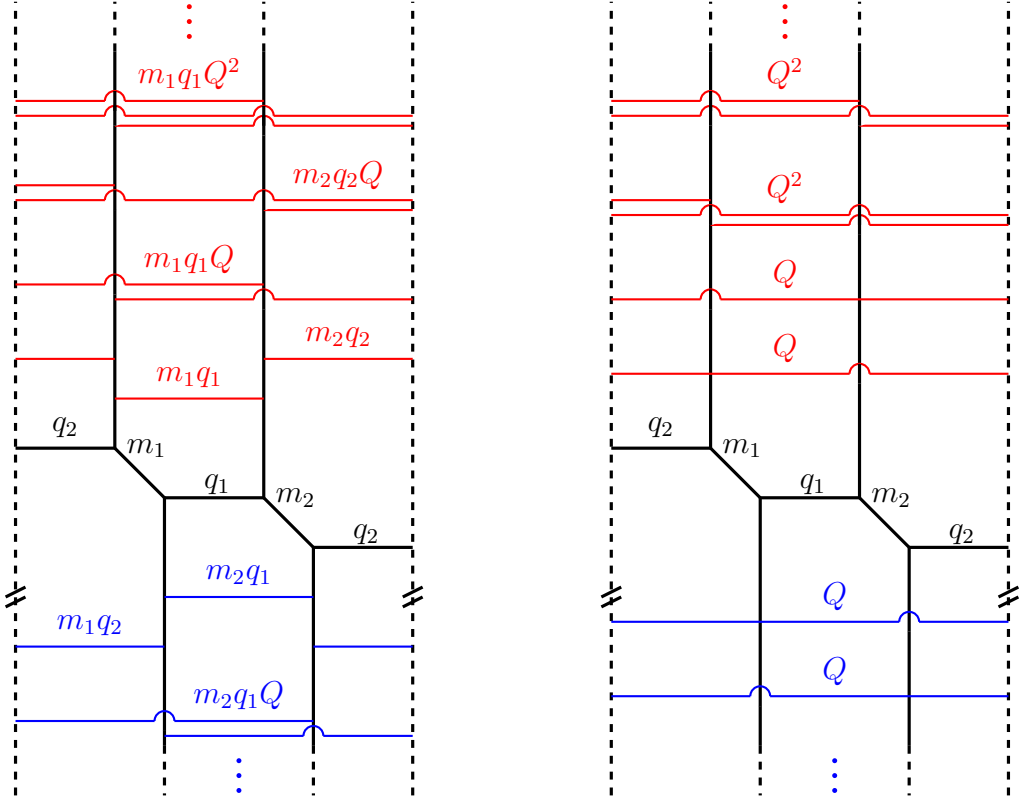

We see that 
\be \label{circp+i} \Z^{\IR^4 \times S^1}_{pert+inst} = PE \left[ \frac
{Q(1-t_1)(1-t_2) -k Q t_1t_2 - kQ + N(q_i,m_i)}
{(1- Q)(1-t_1)(1-t_2)}\right] \ee
with 
\be N(q_i,m_i)=\sum_{i=1}^k( \mt_i t_1t_2+ \tm_{i}^{-1}Q) + \sum_{i=1}^k \sum_{I=1}^{k-1}  (\prod_{s=i}^{i+I-1}\qt_s \tm_s) (\tm_{i+I-1}^{-1}+\tm_{i-1} t_1t_2 - t_1t_2 -\tm_{i+I-1}^{-1}\tm_{i-1})\ee
The formula for $N(q_i,m_i)$ displays $2k^2$ positive terms and $2k(k-1)$ negative terms.

The $2k^2$ positive terms transform in the bifundamental representation of $SU(k) \times SU(k)$, except that they are paired as $x+Q/x$ instead of $x+1/x$. We will see in the next subsection that the correct character of the bifundamental is reproduced once we build the $S^5$ partition function, with a proper analytic continuation. Let us underline that this is a non-trivial check that it is really the $S^5$ partition function which matters for comparison with the M5 brane index.

The $2k(k-1)$ negative terms are all spurious terms to be factored out (Fig.~\ref{spurious-cyl}). Notice that, beacuse of the factor $(1-Q)$ in the denominator, we are really claiming that there is an infinite tower of spurious terms. This fact has a natural interpretetation in the pq-web: for each pair of semi-infinite NS5 branes, we can strech a D1 brane, of length, say, $m_1q_1$, but we can also strech a D1 going around the circle the other way, of length $Q/(m_1q_1)$. As shown graphically in Figure \ref{spurious-cyl}, there are also D1 branes going around the circle more than once, of length $m_1q_1 Q, m_1q_1 Q^2, m_1q_1 Q^3, \ldots$ or $Q^2/(m_1q_1),  Q^3/(m_1q_1), Q^4/(m_1q_1),\ldots$. This explains the $2k(k-1)$ towers of spurious states.

The first part of the numerator in \eqref{circp+i} contains the negative terms $\frac{-k Q t_1t_2 - kQ}{1-Q}$. We interpret these as towers of spurious contributions associated to D1 branes going from one NS5 to itself and wrapping the circle an integer number of times. See the right part of Figure \ref{spurious-cyl}. There are $k$ such contributions, one for every NS5 brane. It looks like these spurious terms are not independent, so to avoid overcounting we have to add back to the partition function a term 
\be -\frac{Q+Qt_1t_2}{1-Q}+t_1t_2 = -\frac{Q+t_1t_2}{1-Q}\ee
We also added the $t_1t_2$ term, which is just a McMahon function, that is $PE[\frac{t_1t_2}{(1-t_1)(1-t_2)}]$, in agreement with \cite{Lockhart:2012vp} for $k=1$.

The final result for $ \Z^{\IR^4 \times S^1}_{pert+inst+spurious}$ for the $k$-nodes circular quiver is
\be \label{circp+i+s}  PE \left[ \frac{Q}{1-Q} + \frac
{- Q - t_1t_2 +  \sum_{i=1}^k( \mt_i t_1t_2+ \tm_{i}^{-1}Q) + \sum_{i=1}^k \sum_{I=1}^{k-1}  (\prod_{s=i}^{i+I-1}\qt_s \tm_s) (\tm_{i+I-1}^{-1}+\tm_{i-1}t_1t_2)}
{(1- Q)(1-t_1)(1-t_2)}\right] \ee

\subsection{Modularity and partition function on $S^5$}
We now patch together $3$ copies of $\Z^{\IR^4 \times S^1}$ to form $\Z^{S^5}$.
First we rewrite $\Z^{\IR^4 \times S^1}$ in terms of modular forms $G_2$ \cite{felder1999},
where for $\text{Im}(\omega_i)>0$
\be
G_2 (z; \omega_0,\omega_1,\omega_2) = PE\left[\frac{-e^{2\pi iz}-e^{2\pi i(-z+\omega_0+\omega_1+\omega_2)}}
                                                  {(1-e^{2\pi i\omega_0})(1-e^{2\pi i\omega_1})(1-e^{2\pi i\omega_2})}\right]
\ee
The exponentiated variables are
$t_1=e^{-\beta\epsilon_1},t_2=e^{-\beta\epsilon_2},m_i=e^{\beta\mu_i},q_i=e^{\beta\tau_i}$
and we defined $Q=\prod_{i=1}^k q_i m_i=e^{\beta\sum_{i=1}^k(\tau_i+\mu_i)}=:e^{\beta\Omega}$.
It's non trivial that our result \eqref{circp+i+s} for $\Z^{\IR^4 \times S^1}_{full}=\Z^{\IR^4 \times S^1}_{pert}\Z^{\IR^4 \times S^1}_{inst}\Z^{\IR^4 \times S^1}_{spurious}$ can be written in terms of $G_2$ functions as
\be
\Z^{\IR^4 \times S^1}_{full}=\frac{G'_2(0;
\frac{\beta\Omega}{2\pi i},\frac{\beta\epsilon_1}{2\pi i},\frac{\beta\epsilon_2}{2\pi i})}
{\eta(\frac{\beta\Omega}{2\pi i})\prod_{i=1}^k \left( G_2(\frac{\beta(\Omega-\mu_i)}{2\pi i};
\frac{\beta\Omega}{2\pi i},\frac{\beta\epsilon_1}{2\pi i},\frac{\beta\epsilon_2}{2\pi i})
\prod_{l=0}^{k-2}G_2(\frac{\beta(\sum_{s=i}^{i+l}(\tau_s+\mu_s)-\mu_{i+l})}{2\pi i};
\frac{\beta\Omega}{2\pi i},\frac{\beta\epsilon_1}{2\pi i},\frac{\beta\epsilon_2}{2\pi i})
\right)}
\ee

The partition function on $S^5$ is obtained taking the product of three copies of $\Z^{\IR^4 \times S^1}_{full}$
\be
\Z^{S^5}=\prod_{\ell=1}^3 \Z^{\IR^4 \times S^1}_{full}(\epsilon_1^{(\ell)},\epsilon_2^{(\ell)},\beta^{(\ell)},\vec q,\vec m)
\ee
These parameters take the following values in the 3 patches of $S^5$  \cite{Nieri:2013vba}
\be
\begin{array}{|c|c|c|c|}
\hline
\ell & \epsilon_1^{(\ell)} & \epsilon_2^{(\ell)} & \beta^{(\ell)} \\ \hline
1 & \omega_2 & \omega_3 & 2\pi i/\omega_1 \\
2 & \omega_3 & \omega_1 & 2\pi i/\omega_2 \\
3 & \omega_1 & \omega_2 & 2\pi i/\omega_3 \\ \hline
\end{array}
\ee
and the double elliptic gamma functions in the $\mathbb{R}^4\times S^1$ partition function are of the form
$G_2(z;\frac{\beta\Omega}{2\pi i},\frac{\beta\epsilon_1}{2\pi i},\frac{\beta\epsilon_2}{2\pi i})$.
They satisfy the modularity property
\be
\begin{aligned}
&G_2\Big(\frac{z}{\omega_1}|\frac{\Omega}{\omega_1},\frac{\omega_2}{\omega_1},\frac{\omega_3}{\omega_1}\Big)
G_2\Big(\frac{z}{\omega_2}|\frac{\Omega}{\omega_2},\frac{\omega_3}{\omega_2},\frac{\omega_1}{\omega_2}\Big)
G_2\Big(\frac{z}{\omega_3}|\frac{\Omega}{\omega_3},\frac{\omega_1}{\omega_3},\frac{\omega_2}{\omega_3}\Big) \\
&=e^{-\frac{\pi i}{12} B_{4,4}(z|\omega_1,\omega_2,\omega_3,\Omega)}
G_2\Big(\frac{z}{\Omega}|\frac{\omega_1}{\Omega},\frac{\omega_2}{\Omega},\frac{\omega_3}{\Omega}\Big)^{-1}
\end{aligned}
\ee
The modular properties of the topological string partition function,
in the case of all the masses equal, have been studied in \cite{Shabbir:2015oxa}.

Using this equation for every triple of $G_2$'s in the $S^5$ partition function we get\footnote{This is computed up to  Bernoulli polynomials and $q^{-1/24}$: harmless terms which do not play any role in our analysis.}
\be\label{ZS5-G2}
\Z^{S^5}=\frac{\prod_{i=1}^k\prod_{l=0}^{k-1}
G_2\Big(\frac{1}{\Omega}\big(\sum_{s=i}^{i+l}(\tau_s+\mu_s)-\mu_{i+l}\big)|-\sigma_1,-\sigma_2,-\sigma_3\Big)}
{\eta(-\sigma_1^{-1})\eta(-\sigma_2^{-1})\eta(-\sigma_3^{-1})G'_2(0|-\sigma_1,-\sigma_2,-\sigma_3)
}
\ee
where $\sigma_i=-\omega_i/\Omega$ 
and the convention $\prod_{i=1}^0\equiv 1$ is used.
Using the modular properties of the $G_r$ \cite{felder1999}
\be
G_r(-z;-\vec\tau)=\frac{1}{G_r(z;\vec\tau)}, \qquad
\eta(\tau^{-1})=\sqrt{i\tau}\eta(-\tau)
\ee
and exponentiated variables $\QI=e^{-2\pi i\omega_i/\Omega}, \tilde q_s=e^{-2\pi i\tau_s/\Omega}, \tilde m_s=e^{-2\pi i\mu_s/\Omega}$
we have\footnote{
$G_2(0)$ contains a zero mode that can be regularized replacing it with
$G'_2(0;\sigma_1,\sigma_2,\sigma_3)=G_2(0;\sigma_1,\sigma_2,\sigma_3)PE[1]=
PE\Big[\frac{-\Qi-\Qii-\Qiii+\Qi\Qii+\Qi\Qiii+\Qii\Qiii-2\Qi\Qii\Qiii}{(1-\Qi)(1-\Qii)(1-\Qiii)}\Big]$.
}
\be
\Z^{S^5}=\frac{G'_2(0|-\sigma_1,-\sigma_2,-\sigma_3)}
{\eta(\sigma_1)\eta(\sigma_2)\eta(\sigma_3)
\prod_{i=1}^k\prod_{l=0}^{k-1}
G_2\big(-(\sum_{s=i}^{i+l}\tau_s \mu_s-\mu_{i+l})/\Omega;\sigma_1,\sigma_2,\sigma_3\big)}
\ee
The variables $\tilde q_s, \tilde m_s$ satisfy
$
\prod_{s=1}^k \tilde q_s \tilde m_s=e^{-2\pi i\frac{1}{\Omega}\sum_{s=1}^k \tau_s+\mu_s}=e^{-2\pi i\frac{1}{\Omega}\Omega}=1
$.
Let us now change variables to $x_1,x_2,\ldots,x_{k-1}$, $y_1,y_2,\ldots,y_{k-1}$,$A$
\be
\tilde m_i = \frac{x_{i}y_{i-1}}{x_{i-1}y_{i}}\frac{A}{\sqrt{\Qi\Qii\Qiii}} \qquad 
\tilde  t_i = \frac{x_{i-1} y_{i-1}}{x_{i}y_{i-2}}\frac{\sqrt{\Qi\Qii\Qiii}}{A}
\ee
and notice that, for $\text{Im}(\sigma_i)>0$, we can rewrite \eqref{ZS5-G2} as
\be
\begin{aligned} \nn
\Z^{S^5}=&PE\Bigg[\frac{1}{\prod_{i=1}^3(1-\QI)}\Big(\Qi\Qii\Qiii-\Qi\Qii-\Qi\Qiii-\Qii\Qiii +\\
&+\sqrt{\Qi\Qii\Qiii}\left[A\left(\sum_{l=1}^k
                              \frac{x_{l+1}}{x_l}\right)
                        \left(\sum_{i=1}^k
                              \frac{y_{i}}{y_{i+1}}\right)
                      +A^{-1}\left(\sum_{l=1}^k
                              \frac{x_{l}}{x_{l+1}}\right)
                             \left(\sum_{i=1}^k
                              \frac{y_{i+1}}{y_{i}}\right)\right]\Big)
                         \Bigg]\end{aligned} \ee
Using the charachters of the (anti)fundamental of $SU(k)$, $\chi^{SU(k)}_{(anti)fund}[x_i]$, our final result is
\be \label{final-res} \begin{aligned}
\Z^{S^5}=& PE\Bigg[\frac{1}{\prod_{i=1}^3(1-\QI)}\Big(\Qi\Qii\Qiii-\Qi\Qii-\Qi\Qiii-\Qii\Qiii +\\
&+\sqrt{\Qi\Qii\Qiii}\left[\chi^{SU(k)}_{fund}[x_i]\chi^{SU(k)}_{antifund}[y_i]\,A  
                      + \chi^{SU(k)}_{antifund}[x_i]\chi^{SU(k)}_{fund}[y_i]\,A^{-1}\right]\Big)
                         \Bigg] .
\end{aligned}
\ee
It is easy to recognize the $6d$ $(1,0)$ superconformal index of a free self-dual tensor \eqref{scitensor} (first line) plus $k^2$ free hypers \eqref{scihyper} (second line).


\section{pq-webs and $q\mathcal{W}$ algebrae}\label{qW}

Five dimensional gauge theories describing the dynamics of the above pq-webs are expected to have a relation with representation
theory of $q\mathcal{W}$ algebrae \cite{Feigin:1995sf,Awata:1995zk}, generalising to five dimensions \cite{Awata:2009ur,Nieri:2013yra} the known AGT relation for four-dimensional class $\mathcal{S}$ theories \cite{Alday:2009aq}.

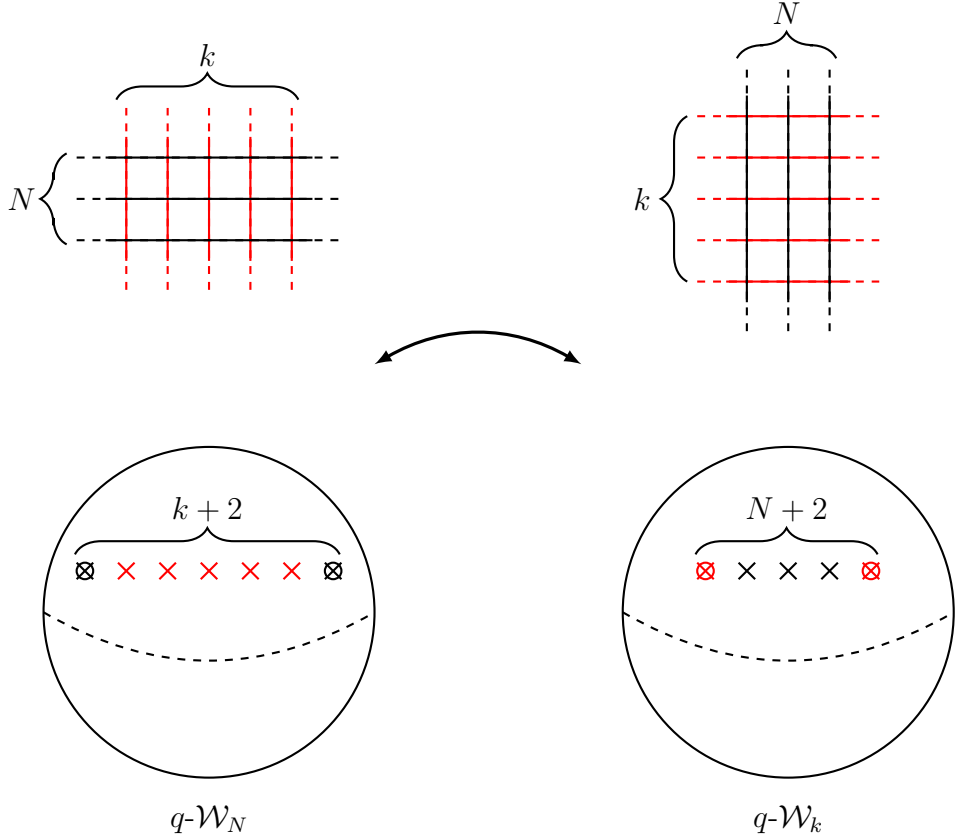
\begin{figure}[h!]
\centering
\begin{tikzpicture}[scale=0.5]

\foreach \x in {3,4,...,7}
{
  \draw[thick, dashed, red]  (\x,13.8) -- (\x,18.2);
  \draw[thick, red]  (\x,14.5) -- (\x,17.5);
}

\foreach \y in {15,16,17}
{
  \draw[thick , dashed]  (1.8,\y) -- (8.2,\y);
  \draw[thick]  (2.5,\y) -- (7.5,\y);
  }

\draw[thick] [decorate,decoration={brace,amplitude=10pt},xshift=-4pt,yshift=0pt]
(2.9,18.4) -- (7.3,18.4) node [black,midway,yshift=0.6cm] 
{$k$};

\draw[thick] [decorate,decoration={brace,amplitude=10pt},xshift=-4pt,yshift=0pt]
(1.7,14.9) -- (1.7,17.1) node [black,midway,xshift=-0.6cm] 
{$N$};

\draw[thick] (5,6) circle (4);
\draw[thick, dashed] (1,6) to [out=-30, in=210] (9,6);

\draw[thick] (2-0.2,7-0.2) -- (2+0.2,7+0.2);  
\draw[thick] (2-0.2,7+0.2) -- (2+0.2,7-0.2);
\draw [thick] (2,7) circle (0.2);

\draw[thick] (8-0.2,7-0.2) -- (8+0.2,7+0.2); 
\draw[thick] (8-0.2,7+0.2) -- (8+0.2,7-0.2);
\draw [thick] (8,7) circle (0.2); 

\foreach \x in {3,4,...,7}
{
\draw[thick, red] (\x-0.2,7-0.2) -- (\x+0.2,7+0.2);  
\draw[thick, red] (\x-0.2,7+0.2) -- (\x+0.2,7-0.2);
}

  \draw[thick] [decorate,decoration={brace,amplitude=10pt},xshift=-4pt,yshift=0pt]
(1.9,7.4) -- (8.3,7.4) node [black,midway,yshift=0.6cm] 
{$k+2$};

\node at (5,1) {$q$-$\mathcal{W}_N$};

\draw[very thick, <->] (9,12) to [out=30, in=150] (14,12);

\foreach \y in {14,15,...,18}
{
  \draw[thick, dashed, red]  (16.8,\y) -- (21.2,\y);
  \draw[thick, red]  (17.5,\y) -- (20.5,\y);
  }

\foreach \x in {18,19,20}
{
  \draw[thick, dashed]  (\x,12.8) -- (\x,19.2);
    \draw[thick]  (\x,13.5) -- (\x,18.5);
}

\draw[thick] [decorate,decoration={brace,amplitude=10pt},xshift=-4pt,yshift=0pt]
(17.9,19.4) -- (20.3,19.4) node [black,midway,yshift=0.6cm] 
{$N$};

\draw[thick] [decorate,decoration={brace,amplitude=10pt},xshift=-4pt,yshift=0pt]
(16.7,14) -- (16.7,18) node [black,midway,xshift=-0.6cm] 
{$k$};

\draw[thick] (19,6) circle (4);
\draw[thick, dashed] (15,6) to [out=-30, in=210] (23,6);

\draw[thick,red] (17-0.2,7-0.2) -- (17+0.2,7+0.2);  
\draw[thick,red] (17-0.2,7+0.2) -- (17+0.2,7-0.2);
\draw [thick,red] (17,7) circle (0.2);

\draw[thick,red] (21-0.2,7-0.2) -- (21+0.2,7+0.2); 
\draw[thick,red] (21-0.2,7+0.2) -- (21+0.2,7-0.2);
\draw [thick,red] (21,7) circle (0.2); 

\foreach \x in {18,19,20}
{
\draw[thick] (\x-0.2,7-0.2) -- (\x+0.2,7+0.2);  
\draw[thick] (\x-0.2,7+0.2) -- (\x+0.2,7-0.2);
}
  
   \draw[thick] [decorate,decoration={brace,amplitude=10pt},xshift=-4pt,yshift=0pt]
(16.9,7.4) -- (21.3,7.4) node [black,midway,yshift=0.6cm] 
{$N+2$};

\node at (19,1) {$q$-$\mathcal{W}_k$};

\end{tikzpicture}
\caption{Linear quiver: a cross is a simple puncture, a cross with a circle is a full puncture.}
\label{fig-CFT-sphere}
\end{figure}

The most interesting consequence of this relation relies in the fact the S-duality in superstring theory, dubbed fiber-base duality in the subclass of topological string amplitudes, 
predicts a duality between $k+2$-point correlators of $q\mathcal{W}_N$ algebrae and $N+2$-point correlators of $q\mathcal{W}_k$ algebrae.
Indeed pq-webs on $\mathbb{R}^2$ are described by five-dimensional gauge theories associated to linear quivers of the kind depicted in Fig.~\ref{fig-CFT-sphere}. 
The brane system on the left-hand side consists in $N$ parallel D4 branes (horizontal black lines) suspended between $k$ NS5 branes (vertical red lines).
As described in Sect.~\ref{inst}, the effective field theory living on the D4 system is a five-dimensional $SU(N)^{k-1}$ linear quiver with $N$-flavors at both ends.
The S-dual system on the right of Fig.~\ref{fig-CFT-sphere} corresponds to a linear quiver $SU(k)^{N-1}$ with $k$ flavors at both the ends. As depicted in Fig.~\ref{fig-CFT-sphere} one expects the 
$S^4\times S^1$ supersymmetric partition function of the first linear quiver to compute the $k+2$-point correlator of $q\mathcal{W}_N$ algebra on the 
sphere with $k$ simple punctures corresponding to semi-degenerate vertex operators of $q$-Toda, and $2$ full $N$-punctures, corresponding to
full vertex operators. Analogously, the $S^4\times S^1$ partition function of the S-dual theory is expected to compute the correlator of $q\mathcal{W}_N$
algebra with $N$ semi-degenerate and two $k$-full insertions (figure \ref{fig-CFT-sphere}).

A first check of this duality is the matching of the dimensions of the space of parameters the correlation functions depend on. Indeed, $k$ simple punctures
on the sphere count $k$ positions of the vertex operator insertions and the corresponding $k$ momenta. On the other hand, the two full $N$-punctures
count each $N-1$ momenta and one position. Overall, taking into account $PSL(2,\mathbb{C})$ symmetry, this amounts to $2(N+k)-3$ parameters.
The counting for the dual correlators is obtained by simply swapping $k$ and $N$.  

A more explicit check can be made in the simplest case $N=1$ by making use of the explicit computations of the supersymmetric partition functions displayed in the previous sections.
In this case the left-hand side of the story reduces to $q$-Heisenberg algebra. A correspondence between this vertex algebra
and five-dimensional gauge theories has been discussed in \cite{Carlsson:2013jka}. According to the duality stated above, the $k+2$-point correlator of $q$-Heisenberg vertex
operators should capture the three-point correlator of $q\mathcal{W}_k$ algebra with two $k$-full and one semi-degenerate insertion \cite{Kozcaz:2010af,Bonelli:2011fq,Mitev:2014isa,Isachenkov:2014eya}.  Studies of the non Abelian cases appeared in \cite{Aganagic:2013tta}.

Let us now proceed to the comparison of the two dual correlators.

We saw in section \ref{linear-quiver} that the partition function
for the $U(1)^{k-1}$ linear quiver theory has three contributions:
perturbative (1-loop), instanton and spurious (due to semi-infinite parallel branes).
The 1-loop and the instanton part can be written as
\be\label{linear-CFT}
\begin{aligned}
\Z^{S^4\times S^1}_{pert+inst}&=PE\bigg[\frac{1}{(1-t_1)(1-t_2)}\Big\{
-(k-1)(t_1+t_2) \\
&-\sum_{i<j}\Big[\big(\frac{x_{i}}{x_{i-1}}\big)\big(\frac{x_{j-1}}{x_{j}}\big)
                +t_1t_2\big(\frac{x_{i-1}}{x_{i}}\big)\big(\frac{x_{j}}{x_{j-1}}\big)
                +\big(\frac{y_{i-1}}{y_{i}}\big)\big(\frac{y_{j}}{y_{j-1}}\big)
                +t_1t_2\big(\frac{y_{i}}{y_{i-1}}\big)\big(\frac{y_{j-1}}{y_{j}}\big)\Big] \\
&+\sqrt{t_1t_2}\Big[
A\big(\sum_{l=1}^k \frac{x_l}{x_{l-1}}\big)(\sum_{i=1}^k \frac{y_{i-1}}{y_{i}}\big)+
A^{-1}\big(\sum_{l=1}^k \frac{x_{l-1}}{x_{l}}\big)(\sum_{i=1}^k \frac{y_{i}}{y_{i-1}}\big)\Big]
\Big\}\bigg]
\end{aligned}
\ee
where we included the 1-loop contribution of the $N-1$ vector multiplets that played no role in the identification of the partition function with the M5 brane SCFT index
\footnote{
The $-1$ term in the $PE$ is needed to remove a zero mode, for $SU(N)$ gauge group it corresponds to the Haar measure on the $S^4\times S^1$ \cite{Kim:2012gu}.
}
\be
\Z^{S^4\times S^1}_\text{pert,vector}=PE\left[\frac{-1-t_1t_2}{(1-t_1)(1-t_2)}+1\right]=PE\left[\frac{-t_1-t_2}{(1-t_1)(1-t_2)}\right].
\ee
By introducing the new variables $\alpha_i,\tilde\alpha_i,\varkappa$ defined by
\be
\begin{aligned}
x_1&=\prod_{i=1}^{k-1}\frac{e^{-\frac{\beta}{k}(\alpha_i-\alpha_{i+1})(i-k)}}{t_1t_2}, \qquad
y_1  =\prod_{i=1}^{k-1}\frac{e^{+\frac{\beta}{k}(\tilde\alpha_i-\tilde\alpha_{i+1})(i-k)}}{t_1t_2},  \\
x_n &=x_1^n\,\prod_{i=1}^{n-1}\frac{e^{-\beta(\alpha_i-\alpha_{i+1})(n-i)}}{t_1t_2}, \qquad
y_n   =y_1^n\,\prod_{i=1}^{n-1}\frac{e^{+\beta(\tilde\alpha_i-\tilde\alpha_{i+1})(n-i)}}{t_1t_2}, \qquad
n=2,\dots,k-1 \\
A^k&=\frac{e^{-\beta\varkappa}}{(t_1t_2)^{k/2}}
\end{aligned}
\ee
we can rewrite \eqref{linear-CFT} in a form which is more suitable for the comparison with the $q \mathcal{W}_k$ correlator
\be\label{q3point}
\begin{aligned}
\Z^{S^4\times S^1}_{pert+inst}(\alpha,\tilde\alpha,\varkappa)=
\frac{\Upsilon'_q(0)^{k-1}
\prod_{e>0}\Upsilon_q(\langle Q-\alpha,e\rangle)\Upsilon_q(\langle Q-\tilde\alpha,e\rangle)}
      {\prod_{i,j=1}^k \Upsilon_q(\frac{\varkappa}{k}+\langle \alpha-Q,h_i\rangle+\langle \tilde\alpha-Q,h_j\rangle)}
\end{aligned}
\ee
where $e$ are the positive roots of the ${ A}_{k-1}$ gauge group,
$Q=(\epsilon_1+\epsilon_2)\rho$, $\rho$ is the Weyl vector (half the sum of all positive roots),
$h_i$ are the weights of the fundamental representation and
$\langle\cdot,\cdot\rangle$ denotes the scalar product on the root space.
The $\Upsilon_q$ function, with $q=e^{-\beta}$ can be defined as follows
\be
\Upsilon_q(x|\epsilon_1,\epsilon_2)=(1-q)^{-\frac{1}{\epsilon_1\epsilon_2}\left(x-\frac{Q}{2}\right)^2}
PE\left[\frac{-q^x-q^{-x}t_1t_2}{(1-t_1)(1-t_2)}\right].
\ee
For $A_{k-1}$ the above can be written in term of $k$-dimensional vectors $u_i$, whose $i$th entry is one and all others zero,
as
\begin{align}
e&=u_i-u_j, \qquad 1\le i<j\le k \\
\rho&=\frac{1}{2}\sum_{i=1}^k(k+1-2i)u_i, \qquad
h_i=-u_i+\frac{1}{k}\sum_{j=1}^k u_j
\end{align}
A $(k-1)$-dimensional vector of fields $\phi$ can be expanded on the base of the simple roots $\hat e_i$ of the $A_{k-1}$
\be
\phi = \sum_{i=1}^{k-1} \phi_i \hat e_i=\sum_{i=1}^{k-1} \phi_i (u_i-u_{i+1}).
\ee

\begin{figure}
\centering
\begin{tikzpicture}[scale=0.30]

\draw[thin,dashed,black!40] (9,2.5) -- (34,2.5);
\draw[thin,dashed,black!40] (4.5,.5) -- (20,.5);
\draw[<->,thick] (17,.5)  -- (17,2.5) node[above] {$\prod_i m_i$};


\draw[thick] (-2,3.5) -- (5,3.5) -- (10,8.5); \draw[thick,dashed] (10,8.5) -- ++(1,1);
\draw[thick] (-2,3) -- (7,3) -- (12,8);\draw[thick,dashed] (12,8) -- ++(1,1);
\draw[thick] (-2,2.5) -- (9,2.5) -- (14,7.5);\draw[thick,dashed] ((14,7.5) -- ++(1,1);
\draw[thick] (-2,.5) -- (4.5,.5) -- (5,1) -- (6.5,1) -- (7,1.5) -- (8.5,1.5) -- (9,2) -- (10.5,2) -- (15.5,7);
\draw[thick,dashed] (15.5,7) -- ++(1,1);
\draw[thick] (5,3.5) -- (5,1); \draw[thick] (7,3) -- (7,1.5); \draw[thick] (9,2.5) -- (9,2); 
\draw[thick] (4.5,.5) -- (4.5,.5-5);\draw[thick,dashed] (4.5,.5-5) -- ++(0,-1.5);
\draw[thick] (6.5,1) -- (6.5,1-5);\draw[thick,dashed] (6.5,1-5) -- ++(0,-1.5);
\draw[thick] (8.5,1.5) -- (8.5,1.5-5);\draw[thick,dashed] (8.5,1.5-5) -- ++(0,-1.5);
\draw[thick] (10.5,2) -- (10.5,2-5);\draw[thick,dashed] (10.5,2-5) -- ++(0,-1.5);
;

\draw[fill=white] (-2,3) circle (0.7);  
\draw (-2-.5,3-.5) -- (-2+.5,3+.5);
\draw (-2-.5,3+.5) -- (-2+.5,3-.5);

\draw[fill=white] (-2,0.5) circle (0.7);  
\draw (-2-.5,0.5-.5) -- (-2+.5,0.5+.5);
\draw (-2-.5,0.5+.5) -- (-2+.5,0.5-.5);

;

\fill [fill=white] (5,3) circle (0.3);   
\draw[thick] (5+0.3,3) arc (0:180:0.3);
\draw[thick] (5,3-0.3) -- (5,3+0.3);

\fill [fill=white] (5,2.5) circle (0.3);   
\draw[thick] (5+0.3,2.5) arc (0:180:0.3);
\draw[thick] (5,2.5-0.3) -- (5,2.5+0.3);

\fill [fill=white] (7,2.5) circle (0.3);   
\draw[thick] (7+0.3,2.5) arc (0:180:0.3);
\draw[thick] (7,2.5-0.3) -- (7,2.5+0.3);


\draw[thick] (21,.5) -- (27.5,.5) -- (28,1) -- (29.5,1) -- (30,1.5) -- (31.5,1.5) -- (32,2) -- (33.5,2) -- (34,2.5) -- (40.5,2.5);

\draw[thick] (27.5,.5) -- (27.5,.5-5);\draw[thick,dashed] (27.5,.5-5) -- ++(0,-1.5);
\draw[thick] (29.5,1) -- (29.5,1-5);\draw[thick,dashed] (29.5,1-5) -- ++(0,-1.5);
\draw[thick] (31.5,1.5) -- (31.5,1.5-5);\draw[thick,dashed] (31.5,1.5-5) -- ++(0,-1.5);
\draw[thick] (33.5,2) -- (33.5,2-5);\draw[thick,dashed] (33.5,2-5) -- ++(0,-1.5);

\draw[thick] (28,1) -- (28,1+5);\draw[thick,dashed] (28,1+5) -- ++(0,1.5);
\draw[thick] (30,1.5) -- (30,1.5+5);\draw[thick,dashed] (30,1.5+5) -- ++(0,1.5);
\draw[thick] (32,2) -- (32,2+5);\draw[thick,dashed] (32,2+5) -- ++(0,1.5);
\draw[thick] (34,2.5) -- (34,2.5+5);\draw[thick,dashed] (34,2.5+5) -- ++(0,1.5);

\draw[fill=white] (21,.5) circle (0.7);  
\draw (21-.5,.5-.5) -- (21+.5,.5+.5);
\draw (21-.5,.5+.5) -- (21+.5,.5-.5);

\draw[fill=white] (40.5,2.5) circle (0.7);  
\draw (40.5-.5,2.5-.5) -- (40.5+.5,2.5+.5);
\draw (40.5-.5,2.5+.5) -- (40.5+.5,2.5-.5);

\end{tikzpicture}

\caption{On the left a generalized pq-web which can be obtained from Higgling the $T_{N=4}$ SCFT \cite{Benini:2009gi}, the global symmetry is $S(U(1) \times U(1)) \times SU(N) \times SU(N)$ (if $N>2$, for $N=2$ it is a standard pq-web with $SU(2)^3$ symmetry). On the right the standard pq-web, one D5 intersecting $k$ NS5's, the global symmetry is the same, $SU(N) \times SU(N) \times U(1)$. The symbols $\otimes$ represent D7 branes. The two pq-webs are related by a sequence of $N$ Hanany-Witten transitions, that, starting from the right, create the $N-1$ D5's and also bend the upper part of the $N$ NS5's. The vertical displacement between the two D7 branes attached to the semi-infinite D5's does not change, and it equals the product of all the masses in the linear quiver $\prod_i m_i=A^k$. The difference between the two partition functions is just that the left diagram has one more spurious term, which is precisely $PE[\frac{-A^{-k}-A^kt_1t_2}{(1-t_1)(1-t_2)}]=\Upsilon_q(\varkappa)$.}\label{fig-TN}
\end{figure}
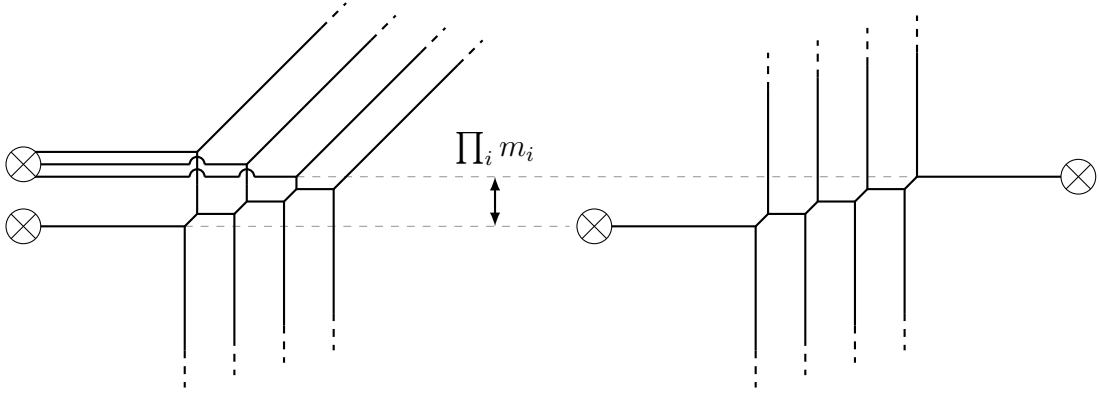

Formula \eqref{q3point}  can be compared to the 
three-point  correlation
function $C_q(\alpha,\tilde\alpha,\varkappa h_{k-1})$ with one degenerate insertion 
(parallel  to the highest weight of antifundamental representation $ h_{k-1}$)
of the $k$-$q$Toda theory with central charge $c=k-1+12\langle Q,Q\rangle$,
that has been conjectured in \cite{Mitev:2014isa,Isachenkov:2014eya}. 
The two formulae are different by a factor $\Upsilon_q(\varkappa)$ in the numerator, which is due to
the fact that the computation in \cite{Mitev:2014isa} corresponds to the generalized pq-web \cite{Benini:2009gi} diagram on left side of Fig.~\ref{fig-TN},
while ours corresponds to the standard pq-web on the right hand side. The two diagrams are related by moving one D7 brane all the way through
the $k$ NS5-branes keeping track of the Hanany-Witten brane creation effect, see \cite{Benini:2009gi,Kim:2014nqa,Kim:2015jba}. 
From the viewpoint of the index computation, on which we focus in this paper, this extra factor is a spurious one due to the contribution
of strings stretching between the two sets of parallel horizontal branes, which are present only in the left-hand side brane diagram.
From the viewpoint of q-deformed CFT, the left-hand side diagram is more natural and has also the correct four-dimensional limit.

\begin{figure}[h!]
\centering
\begin{tikzpicture}[scale=0.5]

\foreach \x in {3,4,...,7}
{
  \draw[thick, dashed, red]  (\x,13.8) -- (\x,18.2);
  \draw[thick, red]  (\x,14.5) -- (\x,17.5);
}
\foreach \y in {15,16,17}
{
  \draw[thick]  (2,\y) -- (8,\y);
  \draw[thick]  (2-.2,\y-.1) -- (2,\y+.1) (8-.2,\y-.1) -- (8,\y+.1);
  \draw[thick]  (2,\y-.1) -- (2+.2,\y+.1) (8,\y-.1) -- (8+.2,\y+.1);
}

\draw[thick] [decorate,decoration={brace,amplitude=10pt},xshift=-4pt,yshift=0pt]
(2.9,18.4) -- (7.3,18.4) node [black,midway,yshift=0.6cm] 
{$k$};

\draw[thick] [decorate,decoration={brace,amplitude=10pt},xshift=-4pt,yshift=0pt]
(1.7,14.9) -- (1.7,17.1) node [black,midway,xshift=-0.6cm] 
{$N$};

\draw[thick] (17,16) ellipse (5 and 3.5);
\draw[thick] (14.5,17.1)  to [out=-30, in=210] (19.5,17.1);
\draw[thick] (16,16.5) to [out=30, in=150] (18,16.5);

\foreach \x in {3,4,...,7}
{
\draw[thick, red] (1\x+2-0.2,14-0.2) -- (1\x+2+0.2,14+0.2);  
\draw[thick, red] (1\x+2-0.2,14+0.2) -- (1\x+2+0.2,14-0.2);
}
  
  \draw[thick] [decorate,decoration={brace,amplitude=10pt},xshift=-4pt,yshift=0pt]
(14.9,14.4) -- (19.3,14.4) node [black,midway,yshift=0.6cm] 
{$k$};

\node at (17,11.5) {$q$-$\mathcal{W}_N$};

\end{tikzpicture}
\caption{Circular quiver: a cross is a simple puncture.}
\label{torus}
\end{figure}
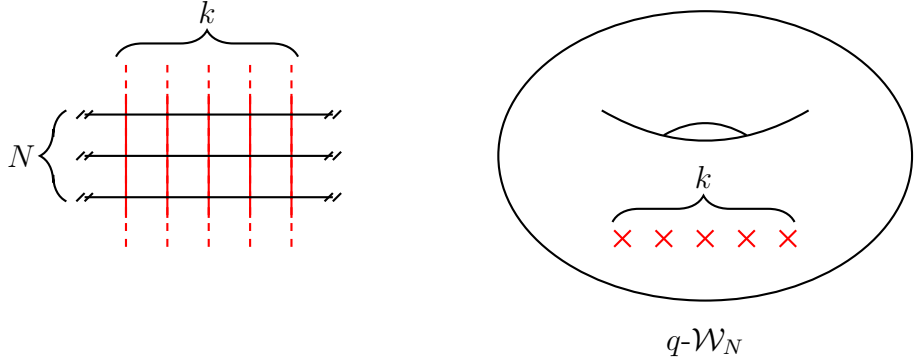

Let us now make some comments on the case of pq-webs on the cylinder. As explained in Sect.~\ref{pqweb-cyl}
this is described in terms of circular quivers. In this case one expects a relation with correlators of $q\mathcal{W}$ algebrae on a torus.
For the case $N=1$ it has indeed been shown in \cite{Carlsson:2013jka} that the one point chiral correlator of $q$-Heisenberg on the torus
computes the $k=1$ circular quiver partition function on $\mathbb{R}^4\times S^1$. The results of Sect.~\ref{circular} should correspond
to the $k$-point chiral correlator of the same vertices, see Fig.~\ref{torus}. It would be interesting to  check this relation explicitly.

\section{Open questions}

There are some open questions which in our view deserve further investigation. Let us briefly mention them:
\begin{itemize}
\item
a natural extension of our work, on which we hope to report soon, is to fully compactify the pq-web brane diagram on the torus. This amounts to compactify 
the D6 branes in Fig.~\ref{fig1} on a circle, and has a link to topological string amplitudes on elliptically fibered Calabi-Yau's 
which are relevant for the classification and study of $\mathcal{N}=(1,0)$ SCFT \cite{Haghighat:2014vxa}.
From the viewpoint of the index computation this set-up would also be useful to analyse the issue
of spurious factors due to the strings stretched between semi-infinite branes, which should appear in the decompactification limit.
From the viewpoint of deformed $\mathcal{W}$-algebrae, the analysis of the S-duality of pq-web diagrams on the torus should be useful
to investigate elliptic $\mathcal{W}$-algebrae as considered in
\cite{Nieri:2015dts,Iqbal:2015fvd,Mironov:2016cyq}.

\item
it would be very interesting to analyze the case of non-abelian theories, which would provide information about interacting
M5 brane systems. This implies the integration over the Coulomb branch parameters and a full control of the polar structure of non-abelian
Nekrasov partition functions in five and six dimensions. We are currently investigating this problem.

\item
the study of S-duality of linear and circular pq-webs and its relation with q$\mathcal{W}$-algebrae has to be further investigated, in 
order to have an independent proof of the plethystic formulae for the supersymmetric partition functions derived in this paper as well as their
interpretation in terms of dualities of q-deformed correlators. In particular it is not clear to us what is the S-dual of the circular quiver displayed in figure \ref{torus}. 

\item
the interpretation of our results in terms of quantum integrable systems is to be analysed. In this context,
it would be useful to study the insertion of defect operators in the supersymmetric partition functions and their brane
realization.

\end{itemize}

\acknowledgments
We thank Can Kozcaz, Constantinos Papageorgakis, Sara Pasquetti and Diego Rodriguez-Gomez for useful discussions.
The research of G.B. is supported by the INFN project ST\&FI.
The research of S.B., M.R. and A.T. is supported by the INFN project GAST.
G.B. and M.R. would like to thank the COST action "The String Theory Universe" MP1210 and the Theoretical Physics Department of the
University of Oviedo for hosting in a very collaborative environment.
\appendix

\section{Nekrasov partition function}\label{nek}
{\small

In this appendix we recall the definition of the Nekrasov partition function
\cite{Nekrasov:2002qd, Flume:2002az, Bruzzo:2002xf}.

The 1-loop part for the circular $U(N)$ quiver with $k$ nodes is
\be\label{1loop}
Z^\text{1-loop}_{U(N)^k}(\{\vec a_i\},\{\mu_i\})
=\prod_{i=1}^k z^\text{1-loop}_{vec}(\vec a_i) z^\text{1-loop}_{bif}(\vec a_i,\vec a_{i+1},\mu_i)
\ee
where
\be\label{z1loop}
\begin{aligned}
z^\text{1-loop}_{vec}(\vec a)&=\prod_{\alpha,\gamma=1}^N \prod_{m,n>1}
  \sinh\frac{\beta}{2}(a_\alpha-a_\gamma+m\epsilon_1+n\epsilon_2) \\
z^\text{1-loop}_{bif}(\vec a,\vec b, \mu)&=\prod_{\alpha,\gamma=1}^N \prod_{m,n>1}
  \sinh\frac{\beta}{2}(a_\alpha-b_\gamma-\mu +m\epsilon_1+n\epsilon_2)^{-1}.
\end{aligned}
\ee

The instanton partition function for the circular $U(N)$ quiver with $k$ nodes is
\be\label{nekr}
Z_{U(N)^K}(\{\vec a_i\},\{\q_i\} , \{ \mu_i\})=\sum_{\{\vec Y_1,\dots \vec Y_k\}}\prod_{i=1}^k \q_i^{|\vec Y_i|} 
z_{vec}(\vec a_i, \vec Y_i)\,z_{bif}(\vec a_i,\vec a_{i+1},\vec Y_i,\vec Y_{i+1},\mu_i)
\ee
where 
$\vec a_{k+1}\equiv \vec a_1$, $\q_{k+1}\equiv \q_1$, $\mu_{k+1}\equiv \mu_1$.
And
\be
\begin{aligned}
z_{vec}(\vec a,\vec Y)=\prod_{\alpha,\gamma=1}^N \prod_{s\in Y_\alpha}
 &\sinh\frac{\beta}{2}\left[a_\alpha-a_\gamma-\epsilon_1 L_{Y_\gamma}(s)+\epsilon_2\left(A_{Y_\alpha}(s)+1\right)\right]^{-1} \\  
 &\sinh\frac{\beta}{2}\left[a_\gamma-a_\alpha+\epsilon_1 \left(L_{Y_\gamma}(s)+1\right)-\epsilon_2 A_{Y_\alpha}(s)\right]^{-1}
\end{aligned}
\ee
\be
\begin{aligned}
z_{bif}(\vec a,\vec b,\vec Y,\vec W,\mu)=\prod_{\alpha,\gamma=1}^N \prod_{s\in Y_\alpha}
 &\sinh\frac{\beta}{2}\left[a_\alpha-b_\gamma-\epsilon_1 L_{W_\gamma}(s)+\epsilon_2\left(A_{Y_\alpha}(s)+1\right)-\mu\right] \\
                                         \prod_{t\in W_\gamma}
 &\sinh\frac{\beta}{2}\left[a_\alpha-b_\gamma+\epsilon_1 \left(L_{Y_\alpha}(t)+1\right)-\epsilon_2 A_{W_\gamma}(t)-\mu\right]
\end{aligned}
\ee

These expressions can be substituted in \eqref{nekr}
with\footnote{
It is easy to check this analytically when $k=2$. We have checked it also for $k=3,4$ with Mathematica up to instanton number $10$.}
\be
\begin{aligned}
z'_{vec}(\vec a,\vec Y)=\prod_{\alpha,\gamma=1}^N\prod_{s\in Y_\alpha}
                          &\left(1-e^{-\beta(a_\alpha-a_\gamma)}t_1^{-L_{Y_\gamma}(s)}t_2^{A_{Y_\alpha}(s)+1}\right)^{-1}\times \\
                          &\left(1-e^{-\beta(a_\gamma-a_\alpha)}t_1^{L_{Y_\gamma}(s)+1}t_2^{-A_{Y_\alpha}(s)}\right)^{-1} \\
z'_{bif}(\vec a,\vec b,\vec Y,\vec W,m)=\prod_{\alpha,\gamma=1}^N\prod_{s\in Y_\alpha}
&\left(1-e^{-\beta(a_\alpha-b_\gamma)}t_1^{-L_{W_\gamma}(s)}
                                      t_2^{A_{Y_\alpha}(s)+1}m\right)\times \\
                          \prod_{t\in W_\gamma}
&\left(1-e^{-\beta(a_\alpha-b_\gamma)}t_1^{L_{Y_\alpha}(t)+1}
                                     t_2^{-A_{W_\gamma}(t)}m\right)
\end{aligned}
\ee
redefining $q_i=\q_i/\sqrt{m_i m_{i+1}}$.
The 5D exponentiated variables are $t_1=e^{-\beta\epsilon_1}$, $t_2=e^{-\beta\epsilon_2}$, $m_i=e^{\beta\mu_i}$
and $q_i=e^{\beta\tau_i}$. 

To consider the instanton contribution of the linear $U(N)$ quiver with $k-1$ nodes it is sufficient to take the limit $q_k\to 0$.
This correspond to freeze the $k$-th gauge group, indeed $q_k=\exp(\beta\tau_k)$ where $\beta=2\pi i R_{S^1}$ and
$\tau=\frac{4\pi i}{g_{YM}^2}$. So $q_k\sim\exp(-1/g_{YM}^2)\to 0$ when $g_{YM}\to 0$.

We obtain a linear quiver gauge theory with $k-1$ $U(N)$ gauge groups, $k-2$ massive bifundamentals and
$2k$ massive flavour at the endpoints of the quiver: $k$ in the fundamental representation at one endpoint
and $k$ in the antifundamental at the other endpoint.
This is because
\be
z_{bif}(\vec a,\vec b,\vec Y ,\vec \varnothing,\mu)=\prod_{\gamma=1}^N z_{fund}(\vec a, \vec Y, \mu+b_\gamma),
\qquad
z_{bif}(\vec a,\vec b,\vec \varnothing, \vec W,\mu)=\prod_{\gamma=1}^N z_{antif}(\vec b, \vec W, \mu-a_\gamma),
\ee
where  
\be
z_{fund}(\vec a, \vec Y, \mu)=\prod_{\alpha=1}^N\prod_{(i,j)\in Y_\alpha}
\sinh\frac{\beta}{2}\left[a_\alpha+i \epsilon_1 +j \epsilon_2-\mu\right]=
-z_{antif}(\vec a, \vec Y, \mu-a_\alpha).
\ee

}

\end{document}